\documentclass[sigconf,nonacm,screen,authorversion]{acmart}

\newcommand\vldbavailabilityurl{\url{https://github.com/wheatman/BYO}}

\usepackage{microtype}
\usepackage[subtle]{savetrees}
\usepackage{xspace}
\usepackage{amsmath}
\usepackage{pifont}
\usepackage{amsthm}
\usepackage{amstext}
\usepackage{amsfonts}
\usepackage{tikz}
\usepackage{cleveref}
\usepackage{pgfplots}
\pgfplotsset{compat=1.9}
\usepackage{algorithm}
\usepackage{cases}
\usepackage{algpseudocode}
\usepackage{wrapfig}
\usepackage{url}
\usepackage{enumerate}
\usepackage{enumitem}
\setitemize{leftmargin=*,noitemsep,topsep=0pt,parsep=0pt,partopsep=0pt}
\usepackage[normalem]{ulem}
\usepackage{relate}
\usepackage{balance}
\usepackage{proc}
\usepackage{float}
\usepackage{flafter}
\usepackage{stfloats}
\usepackage{makecell}
\usepackage{listings}
\usepackage{placeins}
\usepackage[font=normal]{caption}
\usepackage{lipsum}
\usepackage{marginnote}
\usepackage{subcaption}
\usepackage{bbding}
\usepackage{multirow}
\usepackage{multicol}
\usepackage{xcolor,colortbl}
\usepackage{tcolorbox}

\usetikzlibrary{calc,matrix, shapes.multipart,patterns}
\pgfplotsset{
    discard if/.style 2 args={
        x filter/.append code={
            \edef\tempa{\thisrow{#1}}
            \edef\tempb{#2}
            \ifx\tempa\tempb
                
            \fi
        }
    },
    discard if not/.style 2 args={
        x filter/.append code={
            \edef\tempa{\thisrow{#1}}
            \edef\tempb{#2}
            \ifx\tempa\tempb
            \else
                
            \fi
        }
    },
}

\pgfplotsset{compat=1.8,
    /pgfplots/ybar legend/.style={
    /pgfplots/legend image code/.code={%
       \draw[##1,/tikz/.cd,yshift=-0.25em]
        (0cm,0cm) rectangle (3pt,0.8em);},
   },
}

\definecolor{mygreen}{rgb}{0,0.6,0}
\definecolor{mygray}{rgb}{0.5,0.5,0.5}
\definecolor{mymauve}{rgb}{0.58,0,0.82}

\usepackage[font=small,aboveskip=0pt,belowskip=1pt]{caption}
\usepackage[compact]{titlesec}
 \titlespacing*{\section}
 {0pt}{.7ex plus .5ex minus .2ex}{.3ex plus .1ex}
 \titlespacing*{\subsection}
 {0pt}{.5ex plus .5ex minus .2ex}{.3ex plus .1ex}

\definecolor{safe-black}{RGB}{0,0,0}
\definecolor{safe-olive}{RGB}{0,73,73}
\definecolor{safe-teal}{RGB}{0,146,146}
\definecolor{safe-pink}{RGB}{255,109,182}
\definecolor{safe-peach}{RGB}{255,160,110}
\definecolor{safe-plum}{RGB}{73,0,146}
\definecolor{safe-cerulean}{RGB}{0,109,219}
\definecolor{safe-lavender}{RGB}{182,109,255}
\definecolor{safe-sky}{RGB}{109,182,255}
\definecolor{safe-baby}{RGB}{182,219,255}
\definecolor{safe-brick}{RGB}{146,0,0}
\definecolor{safe-brown}{RGB}{146,73,0}
\definecolor{safe-orange}{RGB}{219,209,0}
\definecolor{safe-green}{RGB}{36,255,36}
\definecolor{safe-yellow}{RGB}{255,255,109}

\definecolor{magenta4}{rgb}{0.5625,0,0.5625}
\definecolor{green4}{rgb}{0,0.5625,0}
\definecolor{orange4}{rgb}{0.98,0.31,0.09}

\newcommand{\helen}[1]{{ \textcolor{magenta4}{Helen: {#1}}}}
\newcommand{\todo}[1]{{\textcolor{red}{TODO: {#1}}}}

\newcommand{\defn}[1]{\textit{#1}}

\newcommand{\apiname}{Bring Your Own\xspace}
\newcommand{\apinameshort}{BYO\xspace}
\newcommand{\inlineavgspeedup}{$1.06\times$\xspace}
\newcommand{\minefficientslowdown}{$1.16\times$\xspace}

\newcommand{\avgmaxslowdown}{$1.5\times$\xspace}

\newcommand{\numalgorithms}{10\xspace}
\newcommand{\numgraphs}{10\xspace}
\newcommand{\numtotal}{100\xspace}
\newcommand{\numdatastructures}{27\xspace}

\newcommand{\stdset}{\texttt{std::set}\xspace}
\newcommand{\stdunorderedset}{\texttt{std::unordered\_set}\xspace}
\newcommand{\abslbtree}{\texttt{absl::btree\_set}\xspace}
\newcommand{\abslflathash}{\texttt{absl::flat\_hash\_set}\xspace}
\newcommand{\stdvector}{\texttt{std::vector}\xspace}

\newcommand{\numvertices}{\texttt{num\_vertices}\xspace}
\newcommand{\numedges}{\texttt{num\_edges}\xspace}
\newcommand{\degree}{\texttt{degree}\xspace}
\newcommand{\mapneighbors}{\texttt{map\_neighbors}\xspace}

\newcommand{\setapi}{NeighborSet API\xspace}
\newcommand{\containerapi}{GraphContainer API\xspace}
\newcommand{\vertexsubset}{vertexSubset\xspace}
\newcommand{\edgemap}{edgeMap\xspace}

\newcommand{\roadshort}{RD\xspace}
\newcommand{\ljshort}{LJ\xspace}
\newcommand{\orkutshort}{CO\xspace}
\newcommand{\rmatshort}{RM\xspace}
\newcommand{\ershort}{ER\xspace}
\newcommand{\proteinshort}{PR\xspace}
\newcommand{\twittershort}{TW\xspace}
\newcommand{\papersshort}{PA\xspace}
\newcommand{\friendstershort}{FS\xspace}
\newcommand{\kronshort}{KR\xspace}

\newcommand{\figref}[1]         {Figure~\ref{fig:#1}}

\newcommand{\tabref}[1]        {Table~\ref{tab:#1}}
\newcommand{\secref}[1]         {Section~\ref{sec:#1}}

\newcommand{\secreftwo}[2]      {Sections \ref{sec:#1} and~\ref{sec:#2}}

\newcommand{\rmat}{RMAT\xspace}

\newcommand{\para}[1]{\smallskip\noindent\textbf{#1.}}
\renewcommand{\paragraph}[1]{\para{#1}}

\newcommand{\rev}[1]{{{#1}}}
\newcommand{\mymarginpar}[1]{}
\newcommand{\new}[1]{{{#1}}}

\setcopyright{none}

\begin{document}


\title{BYO: A Unified Framework for Benchmarking\\ Large-Scale Graph Containers}
\settopmatter{authorsperrow=4}

\author{Brian Wheatman}
\affiliation{%
  \institution{Johns Hopkins University}
}
\email{wheatman@cs.jhu.edu}

\author{Xiaojun Dong}
\affiliation{%
  \institution{UC Riverside}
}
\email{xdong038@ucr.edu}

\author{Zheqi Shen}
\affiliation{%
  \institution{UC Riverside}
}
\email{zheqi.shen@email.ucr.edu}

\author{Laxman Dhulipala}
\affiliation{%
  \institution{UMD}
}
\email{laxman@umd.edu}

\author{Jakub Łącki}
\affiliation{%
  \institution{Google Research}
}
\email{jlacki@google.com}

\author{Prashant Pandey
}
\affiliation{%
  \institution{University of Utah}
}
\email{pandey@cs.utah.edu}

\author{Helen Xu}
\affiliation{%
  \institution{Georgia Tech}
}
\email{hxu615@gatech.edu}


\begin{abstract}

A fundamental building block in any graph algorithm is a \emph{graph container} -- a data structure used to represent the graph.
Ideally, a graph container enables efficient access to the underlying graph\new{, has low space usage}, and supports updating the graph efficiently.
In this paper, we conduct an extensive empirical evaluation of graph containers designed to support running algorithms on large graphs. To our knowledge, this is the first \emph{apples-to-apples} comparison of graph containers rather than overall systems, which include confounding factors such as differences in algorithm implementations and infrastructure. 

We measure the running time of 10 highly-optimized algorithms across over 20 different containers and 10 graphs.
Somewhat surprisingly, we find that the average algorithm running time does not differ much across containers, especially those that support dynamic updates.
Specifically, a simple container based on an off-the-shelf B-tree is only $1.22\times$ slower on average than a highly optimized static one.
Moreover, we observe that simplifying a graph-container Application Programming Interface (API) to only  a few simple functions incurs a mere $1.16\times$ slowdown compared to a complete API. Finally, we also measure batch-insert throughput in dynamic-graph containers for a full picture of their performance. 

To perform the benchmarks, we introduce \apinameshort, a unified framework that standardizes evaluations of graph-algorithm performance across different graph containers.
\apinameshort extends the Graph Based Benchmark Suite (Dhulipala et al. 18), a
state-of-the-art graph algorithm benchmark, to easily plug into different dynamic graph containers and enable fair comparisons between them on a large suite of graph algorithms.
While several graph algorithm benchmarks have been developed to date, to the best of our knowledge, \apinameshort is the first system designed to benchmark graph containers.
\end{abstract}


\maketitle


\ifdefempty{\vldbavailabilityurl}{}{
\vspace{.3cm}
\begingroup\small\noindent\raggedright\textbf{Artifact Availability:}\\
The source code, data, and/or other artifacts have been made available at \vldbavailabilityurl.
\endgroup
}

\pagenumbering{arabic}

\section{Introduction}\label{sec:intro}

A fundamental design decision in the process of developing any graph algorithm
is the choice of the graph container, that is the data structure that represents
the graph.  This decision can greatly affect both the running time as well as
the space usage of both the graph container and the entire algorithm.  A classic
example used in algorithms textbooks is the difference between an
\emph{adjacency matrix} and an \emph{adjacency list}.  The former is simply a
matrix which uses $n \times n$ bits to store adjacency between all pairs of $n$
graph vertices.  While it supports very efficient queries about the existence of
any edge, its space usage is often prohibitive, especially when applied to
sparse graphs where the number of edges $m$ is close to the number of vertices
$n$.  On the other hand, adjacency list, while much more space efficient, can
require up to $\Theta(n)$ time to determine the existence of any edge.

In practice, it turns out that algorithms typically do not need to query for the
existence of a given edge, and thus the adjacency list idea is more commonly
used in practice.  Specifically, the usual format of choice is the {\em
  Compressed Sparse Row (CSR)} format~\cite{TinneyWa67}, which is an array-based
version of the adjacency list format.  CSR uses an array $A$ of $m$ neighbor ids
to store $m$ edges.  The neighbors of each vertex $v$ form a contiguous fragment
of $A$, and so for each vertex $v$ CSR format additionally stores where this
fragment of $A$ is located.  As a result, CSR requires $O(m + n)$ space to
represent a graph of $n$ vertices and $m$ edges.  Due to its simplicity and good
spatial locality, CSR allows accessing the graph very efficiently.  However,
updating CSR is prohibitively expensive, as inserting a single edge can require
$\Theta(m)$ time due to the fact that all edges are stored in a single flat
array.

To address this limitation, a significant research effort over the past decade
has centered on efficient dynamic-graph containers and their corresponding
systems~\cite{EdgigerMcRi12, KyrolaBlGu12, DeLeoBo21, MackoMaMa15,
  DhulipalaBlSh19,PandeyWhXu21,VanPrWi22,WheatmanBu21,
  WheatmanXu18,WheatmanXu21,DhulipalaBlGu22}. A \defn{dynamic-graph system} is
made up of two parts: the \defn{container} and the \defn{programming framework}.
The container stores the graph topology and handles changes to the graph,
while the programming framework uses an Application Programming Interface (API),
or a specification for how two system components communicate with each other,
provided by the container to express and perform analytics.

Despite the impressive body of existing work on dynamic-graph systems and
containers, at present it is essentially impossible to answer the very basic
question of {\em which} container is appropriate for a given graph application.
A major reason for this situation is that most if not all papers introducing new
dynamic graph containers perform {\em end-to-end} comparisons with existing
systems.  As a concrete example, prior evaluations of the dynamic-graph systems
SSTGraph~\cite{WheatmanBu21} and CPAM~\cite{DhulipalaBlGu22} compare with
earlier dynamic-graph systems such as Aspen~\cite{DhulipalaBlSh19}, but change
not only the container but also important graph-algorithm details, making the
source of any measured improvements unclear.  A second question of no less
importance that is unanswered by existing work is how much performance can be
gained by using very simple ``off-the-shelf'' data structures (e.g., those from
the standard library) to build dynamic graph systems.


In this paper, we introduce \emph{\apiname} (\apinameshort), a unified
programming framework for benchmarking and evaluating graph containers.
%
We use \apinameshort to perform a {\em comprehensive and fair} benchmark of 27
different graph containers, which include both state-of-the-art data structures
such as CPAM~\cite{DhulipalaBlGu22} and SSTGraph~\cite{WheatmanBu21}, as well as
off-the-shelf data structure libraries such as those from the \texttt{std}
standard library and Abseil~\cite{absl}, an open-source standard library from
Google.  These generic data-structure libraries provide a reference
implementation and demonstrate how much performance is left on the table with
simple structures and minimal programming effort.  Our benchmark involves
running 10 fundamental graph algorithms on 10 large graph datasets with up to
$4.2$B edges.

  \begin{figure}[t]
  \centering
  \includegraphics[width=.6\linewidth]{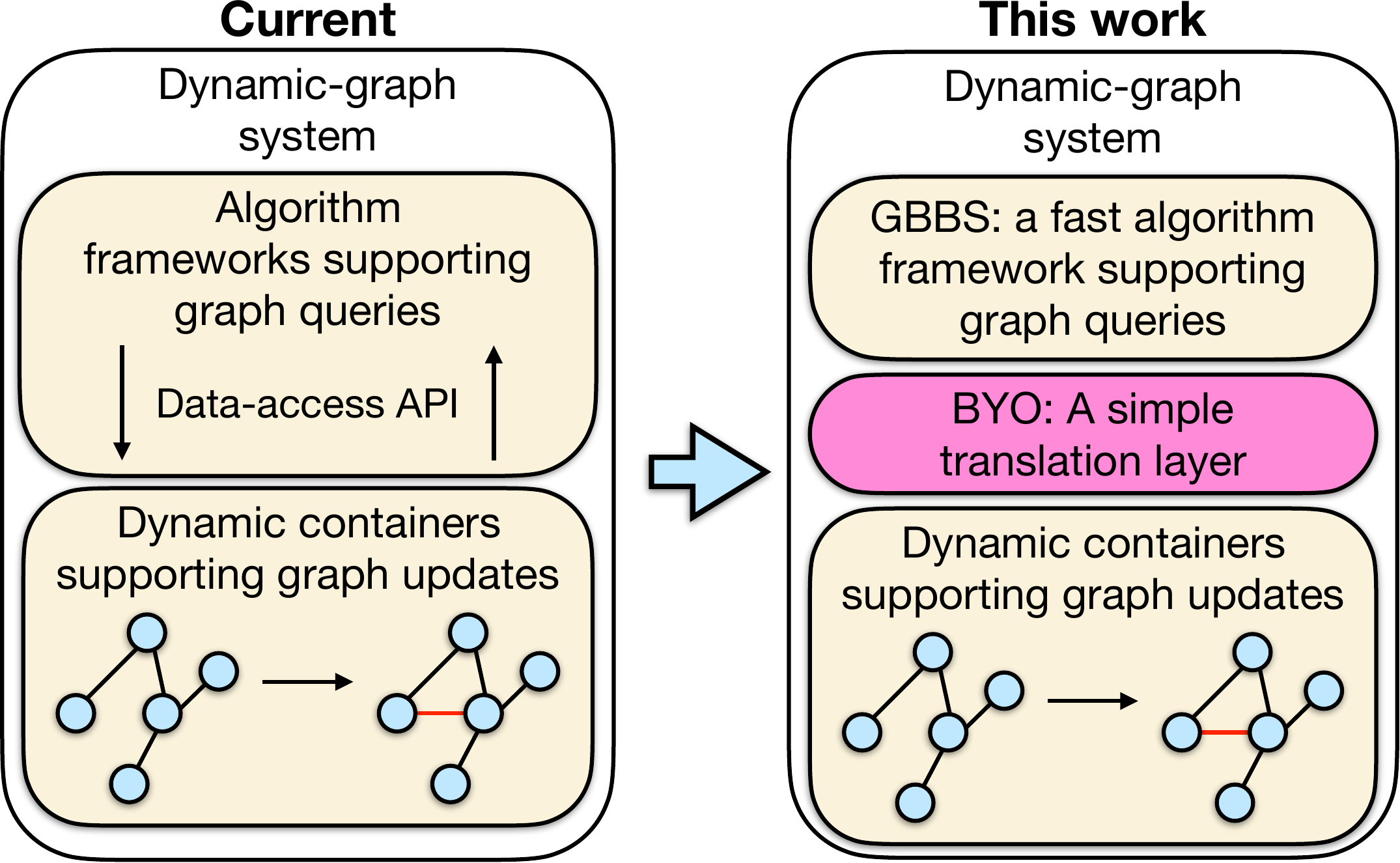}
  \caption{Relationship between \apinameshort, graph-algorithm frameworks, and
    dynamic-graph containers.}
    \label{fig:high-level-api}
  \end{figure}

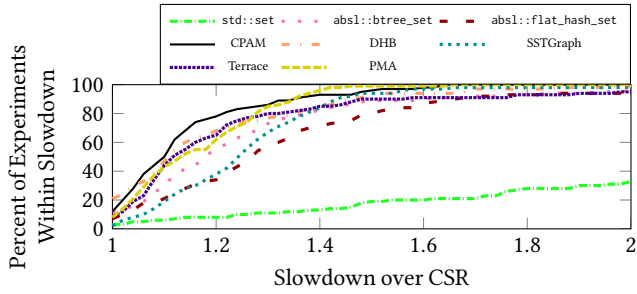
\begin{figure}
    \centering
\begin{tikzpicture}
\begin{axis}[
  xlabel= Slowdown over CSR,
  ylabel style={align=center},
  ylabel= Percent of Experiments\\ Within Slowdown,
  ymax = 100,
  ymin = 0,
  xmax=2,
  xmin=1,
  legend columns=3,
  legend style={at={(1,1)},anchor=south east},
  legend style={font=\tiny, row sep=0pt},
  width=\linewidth, height=3.5cm,
         ]
\addplot +[mark=none, densely dashdotted, very thick, safe-green] table [x=slowdown, y=run_std_set_inplace, col sep=tab] {tsvs/how_close_structures.tsv};
\addlegendentry{\stdset}
\addplot +[mark=none, dash pattern=on \pgflinewidth off 5pt, very thick, safe-pink] table [x=slowdown, y=run_absl_btree_set_inplace, col sep=tab] {tsvs/how_close_structures.tsv};
\addlegendentry{\abslbtree}
\addplot +[mark=none, dash pattern=on 3pt off 7pt, very thick, safe-brick] table [x=slowdown, y=run_absl_flat_hash_set_inplace, col sep=tab] {tsvs/how_close_structures.tsv};
\addlegendentry{\abslflathash}
\addplot +[mark=none, solid, safe-black, thick] table [x=slowdown, y=run_vector_cpam_inplace, col sep=tab] {tsvs/how_close_structures.tsv};
\addlegendentry{CPAM}
\addplot +[mark=none, loosely dashdotted, very thick, safe-peach ] table [x=slowdown, y=run_dhb, col sep=tab] {tsvs/how_close_structures.tsv};
\addlegendentry{DHB}
\addplot +[mark=none, dotted, very thick, safe-teal] table [x=slowdown, y=run_sstgraph, col sep=tab] {tsvs/how_close_structures.tsv};
\addlegendentry{SSTGraph}
\addplot +[mark=none, dash pattern=on \pgflinewidth off .5pt, very thick, safe-plum] table [x=slowdown, y=run_terrace, col sep=tab] {tsvs/how_close_structures.tsv};
\addlegendentry{Terrace}
\addplot +[mark=none, dash pattern=on 3pt off 1pt, very thick, safe-orange] table [x=slowdown, y=run_single_pma, col sep=tab] {tsvs/how_close_structures.tsv};
\addlegendentry{PMA}
\end{axis}
\end{tikzpicture}
    \caption{
    Slowdown of each container compared to CSR.
    Each point ($x, y$) for a given graph container\protect\footnotemark means that the container was at most $x$ times slower than CSR on $y\%$ of experiments.
   A line going up faster implies that the container achieves closer performance relative to CSR on more experiments.  We find that almost all structures are able to perform the majority of the experiments with at most a $1.4\times$ slowdown over CSR. \stdset and absl::* are off-the-shelf containers, while the others are optimized.  Details on the graph containers can be found in Section \ref{sec:datastr}.}
   \label{fig:how_close_structures}
\end{figure}

\paragraph{Fairness of graph-container evaluation}
An interesting, and perhaps surprising finding of our benchmark is the fact the
algorithm performance does not vary drastically between the different containers
when averaging over all graphs and algorithms (see
\figref{how_close_structures}).  This relatively small performance difference
between different graph containers makes them particularly challenging to
benchmark, because a reported performance gain in a proposed dynamic-graph
system may be the result of many factors: the container might be better, the
system may have better algorithm implementations, or the system may use a better
language, compiler, or parallelization library (e.g.,
pthreads~\cite{NicholsBuFa96}, OpenMP~\cite{DagumMe98},
Cilk~\cite{BlumofeJoKu95}, etc.).  We note that varying all of these components
can lead to performance variations which are at least as large in magnitude as
the performance differences we observe between many of the graph containers that
we compare.  Hence, to truly evaluate two graph containers in an
apples-to-apples way, \apinameshort ensures that the framework and all other
infrastructure (i.e., parallelization library, language, compiler) is consistent
across all benchmarks.  While this is a seemingly natural requirement, it was
not fully met in existing papers evaluating graph systems.

\paragraph{Simplified graph-container evaluation}
\apinameshort is based on the Graph Based Benchmark Suite
(GBBS)~\cite{DhulipalaShTs20, DhulBlSh21}, a high-performance graph-algorithm
framework implemented on top of a CSR container.  \apinameshort provides a
minimal translation layer between GBBS and graph containers (e.g., Aspen,
SSTGraph, etc.).  In other words, \apinameshort introduces a simple and abstract
container API, i.e., the API that the containers need to implement, and
implements the popular Ligra/GBBS interface\footnote{GBBS is an iteration of
  Ligra with a richer interface and more algorithm implementations.} using this
API.  This enables users to easily bring their own graph container and connect
it to the programming framework (it suffices if the container implements
\apinameshort's container API), as well as to study their own new algorithms (as
long as they are expressed in the Ligra/GBBS interface).
\mymarginpar{R1D1}\new{\apinameshort is able to represent
  directed, undirected, weighted and unweighted
  graphs.}~\figref{high-level-api} illustrates the relationship between
\apinameshort and other parts of a dynamic-graph system.

We note that the graph container API defined by \apinameshort is very simple.
Specifically, we find that to implement a wide variety of the primitives in
GBBS, \emph{all the graph container developer needs to implement is the map
  primitive} (excluding basic query functions such as \numvertices or
\numedges). \defn{Map} is a functional primitive that applies an arbitrary
function \texttt{f} over a collection of elements. As we shall see, setting
different functions in a map can express other functionality such as reduce and
count. A map can easily be implemented with basic iterators such as those in the
\texttt{C++} standard template library (STL)~\cite{MusserDeSa01, Josuttis12} by
applying the function \texttt{f} to each element in turn.  This feature of
\apinameshort greatly simplifies the process of including a new graph container
in the benchmark.

For comparison, the graph container API (that the container must support) from
GBBS defines 10 primitive neighborhood operations (e.g., map, reduce, scan,
etc.). Similarly, the GraphBLAS specification~\cite{gb-userguide} includes 12
operations (e.g., mxm, assign, apply, etc.) for representing graph algorithms.

\mymarginpar{R1D3, R5D1} \new{
The main technical challenges in \apinameshort were  1) identifying the correct minimal APIs that can generalize to large classes of graph containers and algorithms, 2)  identifying all the code in the original GBBS implementation that makes assumptions about the underlying container and converting them to use modern C++ features that can determine which container functionality to use at compile-time to maximize performance, and 3) simplifying the design to make the translation smooth from the container-developer’s point of view.}

\mymarginpar{R5W3}\new{We built \apinameshort based on GBBS because GBBS has
  been shown to support a wide variety of theoretically and practically
  efficient graph algorithms with better performance than alternatives. As we
  will show in~\secref{eval}, we verify these results and show that GBBS
  achieves $1.06-4.44\times$ speedup on average compared to other frameworks
  (e.g., Ligra~\cite{ShunBl13} and GraphBLAS~\cite{KepnerAaBa16, BulucMaMc17,
    Davis19, Davis23}).
}

\paragraph{Benchmark results}
We perform a cross-cutting evaluation of graph containers and frameworks
along several distinct axes.

The first studies the impact of the graph API when fixing the underlying
container.  In particular, we study performance when moving from a very simple
API (e.g., only supporting \texttt{map}) to a rich API supporting sophisticated
traversal primitives.  Our main finding is \apinameshort using a few very basic
primitives (e.g., map, degree, and the number of edges) is only 1.16$\times$
slower on average than \apinameshort using the full API. However, more advanced
primitives, e.g., parallel map, are necessary to achieve the best performance on
specific instances such as on skewed graphs.

The second axis evaluates existing graph-algorithm frameworks compared to
\apinameshort to ensure that \apinameshort is a good starting point for a
large-scale evaluation and achieves high performance on a variety of
algorithms. We find that on average, \apinameshort achieves competitive
performance on graph algorithms when compared to Ligra~\cite{ShunBl13},
GraphBLAS~\cite{KepnerAaBa16, BulucMaMc17, Davis19, Davis23}, and
GBBS~\cite{DhulipalaShTs20, DhulBlSh21}, which are high-performance
state-of-the-art frameworks.

The third axis studies the impact of different containers by fixing the
algorithm using the \apinameshort API and varying the container.
Our results here have interesting ramifications for the future design of dynamic
graph containers, as well as users of dynamic graph containers. To highlight
just one example, we find that users wishing to use simple ``off-the-shelf''
(i.e., not tailor-made) data structures can build dynamic graph containers using
Abseil B-trees while only incurring a 1.22$\times$ slowdown on average across
all algorithms and graphs over the best static graph container (CSR). However,
off-the-shelf data structures suffer more than specialized data structures in
the worst case on certain problem instances. For example, as shown
in~\figref{how_close_structures}, the Abseil B-tree incurs more slowdown on more
problem settings relative to CPAM~\cite{DhulipalaBlGu22}, a specialized data
structure.  \footnotetext{The set data structures (\stdset, \abslbtree,
  \abslflathash, and CPAM) use the inline optimization described
  in~\secref{byo-api}.}

\apinameshort addresses previous evaluation issues due to different framework
implementations by making sensible optimizations for graph-algorithm performance
accessible to all containers that use \apinameshort.
%
For example, the authors of the SSTGraph graph container implemented the Ligra framework on
top of SSTGraph~\cite{WheatmanBu21} to compare with
Aspen~\cite{DhulipalaBlSh19}, which also implements Ligra. However, the Ligra
implementation in SSTGraph contains additional optimizations for certain
algorithms that enable the overall system to achieve better performance on
certain workloads.  Specifically, SSTGraph found that one of these optimizations
helped by 20\% on Pagerank and 6\% on Connected
Components~\cite{WheatmanBu21}. These optimizations are localized in the
programming framework and could theoretically be applied to any dynamic-graph
container; by incorporating them, we believe that \apinameshort is the first
system that can fairly and reliably isolate performance improvements to the
graph container.

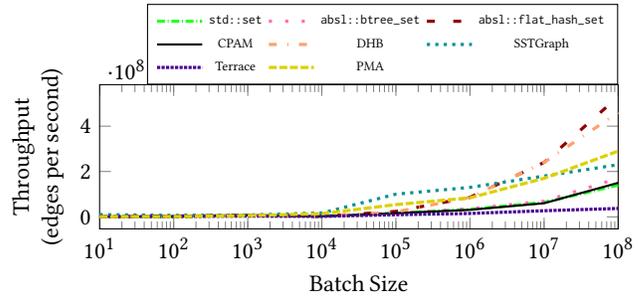
\begin{figure}
    \centering
\begin{tikzpicture}
\begin{axis}[
  xlabel= Batch Size,
  ylabel style={align=center},
  ylabel= Throughput \\ (edges per second),
  xmin=10,
  xmax=100000000,
  legend columns=3,
  legend style={at={(1,1)},anchor=south east},
    legend style={font=\tiny, row sep=0pt},
  xmode=log,
  width=\linewidth, height=3.5cm,
  axis line style=black
         ]
\addplot +[mark=none, densely dashdotted, very thick, safe-green] table [x=batch_size, y=run_std_set_inplace, col sep=tab] {tsvs/insert_with_sort.tsv};
\addlegendentry{\stdset}
\addplot +[mark=none, dash pattern=on \pgflinewidth off 5pt, very thick, safe-pink] table [x=batch_size, y=run_absl_btree_set_inplace, col sep=tab] {tsvs/insert_with_sort.tsv};
\addlegendentry{\abslbtree}
\addplot +[mark=none, dash pattern=on 3pt off 7pt, very thick, safe-brick] table [x=batch_size, y=run_absl_flat_hash_set_inplace, col sep=tab] {tsvs/insert_with_sort.tsv};
\addlegendentry{\abslflathash}
\addplot +[mark=none, solid, safe-black, thick] table [x=batch_size, y=run_vector_cpam_inplace, col sep=tab] {tsvs/insert_with_sort.tsv};
\addlegendentry{CPAM}
\addplot +[mark=none, loosely dashdotted, very thick, safe-peach ] table [x=batch_size, y=run_dhb, col sep=tab] {tsvs/insert_with_sort.tsv};
\addlegendentry{DHB}
\addplot +[mark=none, dotted, very thick, safe-teal] table [x=batch_size, y=run_sstgraph, col sep=tab] {tsvs/insert_with_sort.tsv};
\addlegendentry{SSTGraph}
\addplot +[mark=none, dash pattern=on \pgflinewidth off .5pt, very thick, safe-plum] table [x=batch_size, y=run_terrace, col sep=tab] {tsvs/insert_with_sort.tsv};
\addlegendentry{Terrace}
\addplot +[mark=none, dash pattern=on 3pt off 1pt, very thick, safe-orange] table [x=batch_size, y=run_single_pma, col sep=tab] {tsvs/insert_with_sort.tsv};
\addlegendentry{PMA}
\end{axis}
\end{tikzpicture}
    \caption{\rev{The throughput of inserts for different batch sizes. Of the data structures in this plot, \stdset and absl::*, are off-the-shelf containers, while the others are optimized. Details on the different graph containers can be found in Section \ref{sec:datastr}.}}
    \label{fig:inserts_with_sort}
\end{figure}

The fourth axis evaluates the performance of the dynamic graph containers
when performing batch edge
insertions and deletions.
\apinameshort also integrates numerous off-the-shelf containers
(e.g., Abseil flat hash sets and B-trees), providing a more nuanced picture of
dynamic graph containers built using standard data structures that to the best
of our knowledge is absent in prior evaluations (see
Figure~\ref{fig:inserts_with_sort}).


\section{Related work}

\paragraph{Graph-algorithm benchmarks and frameworks}
Many graph-algorithm frameworks have appeared in the literature, but they have
focused on graph-algorithm performance rather than containers. For example,
several static-graph-algorithm frameworks such as Ligra~\cite{ShunBl13},
GraphBLAS~\cite{KepnerAaBa16, BulucMaMc17, Davis19, Davis23},
Galois~\cite{KulkBuCa09, galois}, and GBBS~\cite{DhulipalaShTs20, DhulBlSh21}
deliver state-of-the-art running times, but are limited to using CSR as the
underlying graph representation.

The GAP benchmark suite~\cite{BeamAsPa15} (and other benchmark suites such as
LDBC graphalytics~\cite{IosupHeNg16}) has had great success in standardizing
evaluations for graph algorithms.  For example, it has been used to
benchmark~\cite{AzadAzBe20} several static-graph-algorithm frameworks including
GraphBLAS as well as DSLs like GraphIt~\cite{ZhangYaBa18}.  However, GAP was not
designed to benchmark dynamic \emph{containers} in a general way.  Specifically,
the GAP specification does not describe what a graph container API should look
like.  Moreover, the reference implementations in GAP are tightly knit with a
CSR graph container.

Further, there are lines of research focused on developing incremental
~\cite{ChengHoKy12,MurrayMcIs13,JiangXuYi21,McShMuIs13,ShiCuSh16,McSherryLaSc20}
and dynamic~\cite{SenSuZh16, VoraGuXu17, MariVo19, MariChVo21, AfarinGaRa23,
  ChenGuZh22, IyerPuPa21, FengMaLi21} algorithms, or designing highly efficient
graph containers \cite{WheatmanXu18,WheatmanBu21, PandeyWhXu21, WheatmanXu21,
  WheatmanBuBu23, DhulipalaBlSh19, DhulipalaBlGu22,
  VanPrWi22,IslamDa23,EdgigerMcRi12} with supports of dynamic updates to the
graphs.  However, all of these works either create an ad-hoc framework with
specialized optimizations, or implement only a few algorithms and compare to a
few of the recent works.  This leads to comparisons between entire systems and
not just between graph containers.

\paragraph{Importance of benchmarking graph containers alone}

\mymarginpar{R4W1}\mymarginpar{R4D1}\mymarginpar{R5W2}\mymarginpar{R5D2}
\new{Significant research effort has been devoted to developing and benchmarking
  graph containers and their corresponding systems. These works have reported
  significant speedups:
\begin{itemize}
\item SSTGraph finds a $1.6\times$ speedup over Aspen~\cite{WheatmanBu21}.
\item Terrace finds a $1.7-2.6\times$ speedup over Aspen~\cite{PandeyWhXu21}.
\item Aspen finds $1.8-15\times$ speedup over prior dynamic data
  structures~\cite{DhulipalaBlSh19}.
    \item VCSR finds as $1.2-2\times$ speedup over PCSR~\cite{9826086}.
    \item PPCSR finds a $1.6\times$ speedup over Aspen~\cite{WheatmanXu21}.
    \item CompressGraph finds a $2\times$ speedup over
      Ligra+~\cite{10.1145/3588684}
    \item Teseo finds frequent speedups of at least $1.5\times$ over other graph
      containers~\cite{DeLeoBo21}
\end{itemize}
However, it is likely that much of the improvements seen in these works are from
factors other than the graph container itself.  }

\paragraph{Pros and  cons of different graph containers}\mymarginpar{R4D5}
\new{Aside from raw performance comparisons, graph containers may support different
functionality and storage guarantees. A few major distinctions include:
\begin{itemize}
\item \textbf{Compressed vs uncompressed.} CPMA~\cite{WheatmanBuBu23},
  CPAM~\cite{DhulipalaBlGu22}, and Aspen~\cite{DhulipalaBlSh19} all support
  compression of the graph, which has been shown in prior work to help
  performance and space usage. In contrast, systems like
  Terrace~\cite{PandeyWhXu21} are uncompressed and use significantly more
  space, which may not work on much larger graphs.
\item \textbf{Functional vs in-place.} Tree-based containers such as CPAM and
  Aspen may support functional updates, which enable concurrent queries and
  batch updates on the graph. Most containers (e.g., Terrace, PPCSR, etc.) store
  the graph in place, so queries and updates must happen in a phased manner.
\item \textbf{Batch updates vs concurrent updates.} Some graph containers (e.g.,
  CPMA, CPAM, and Aspen) provide native support for batch updates, which apply a
  set of updates to the container as one operation. Batch updates improve
  update throughput at the cost of latency compared to concurrent updates, where
  the updates may happen simultaneously, but each individual update is atomic.
\end{itemize}

The focus of this paper is on performance alone, so some of these
differences (e.g., update type or functional storage) are not reflected in the
results.  }



\section{Background}\label{sec:prelim}

This section provides background on graphs and their representations necessary
to understand the data structures studied in this paper. It also reviews the
GBBS framework and describes how it uses different API components to express
graph algorithms. 
Throughout this paper, we use the standard shared-memory model
of parallelism in which we have a set of threads that access a shared memory.


\paragraph{Graphs}
A \defn{graph} is a way of storing objects as \defn{vertices} and connections
between those objects as \defn{edges}.
For simplicity we focus on \emph{unweighted} graphs.
Formally, an unweighted graph $G = (V, E)$ is a set of vertices $V$ and a set of
edges $E$. We denote the number of vertices $n = |V|$ and the number of edges
$m = |E|$.  Each vertex $v \in V$ is represented by a unique non-negative
integer less than $|V|$ (i.e.  $v \in \{0, 1, \ldots, |V| - 1\}$). Each edge is
a 2-tuple $(u, v)$ where $u, v \in V$. For each edge, we refer to the vertex $u$
as the \defn{source} and the vertex $v$ as the \defn{destination}. For
undirected graphs, for each edge $(u, v)$ there exists an edge $(v, u)$. In the
undirected case, the \defn{neighbors} of a vertex $u$ are all vertices $v$ such
that there exists an edge $(u, v)$.  \new{For a weighted graph each edge is a
  3-tuple $(u, v, w)$.} Finally, the \defn{degree} of a vertex in a
graph is the number of neighbors it has.


\subsection{Graph representations}

A graph whose vertex set is $\{0, \ldots, |V|-1\}$ can be thought of and
represented as a sequence $s_1, \ldots, s_{|V|-1}$ of \defn{neighbor sets},
where a neighbor set $s_i$ stores all of the neighbors of a vertex $i$.

\paragraph{Storing graphs with set containers}
The sequence of sets abstraction leads to a classical design for a graph storage
format: a list of pointers (one for each vertex) to pre-selected data structures
holding each vertex’s neighbors. Graph-container developers can trade off
performance properties (e.g., insert vs scan) based on the choice of per-vertex
data structures.

A \defn{set container} is any data structure that stores a unique collection of
elements and can be used to store vertex neighbors. Examples of set containers
include red-black trees~\cite{CLRS}, B-trees~\cite{BayerMc72}, hash
maps~\cite{CLRS}, and arrays. For the purposes of this paper, we define a set
container to support the following operations:

\begin{itemize}
\item \texttt{insert(e)/remove(e)}: Insert/delete element \texttt{e} into/from
  the set.
\item \texttt{map(f)}: Apply the function \texttt{f} to all elements of the set
  (can be implemented with \texttt{C++} style iterators).
\item \texttt{size()}: Return the number of elements in the set.
\end{itemize}

These functionalities are naturally expected from any set container data
structure. For example, the \texttt{C++} standard template library
(STL)~\cite{MusserDeSa01, Josuttis12} includes these functions (as well as
others) in their container specification.

\paragraph{Compressed graphs}
To handle increasingly large graphs, several graph-processing
systems~\cite{ShunGhBl15, DhulBlSh21, DhulipalaBlSh19, DhulipalaBlGu22} include
support for compressed graph formats. Specifically, they provide support for
graphs where neighbor lists are encoded using byte codes~\cite{BlandBlKa04,
  BlandBlKa03} and a parallel generalization~\cite{ShunGhBl15} of byte codes.
Byte codes store a vertex’s neighbor list by difference
encoding~\cite{smith1997scientist} consecutive vertices, with the first vertex
difference encoded with respect to the source. Compression enables larger graphs
to fit in memory and reduces memory traffic, which may help during parallel
processing~\cite{ShunGhBl15, DhulBlSh21, DhulipalaBlSh19, DhulipalaBlGu22}.

All of the {\em compressed} graph representations in this paper use difference
encoding and byte codes to compress the graphs.

\subsection{GBBS API}\label{sec:gbbs-api}
GBBS~\cite{DhulBlSh21} is a shared-memory graph processing framework based on Ligra~\cite{ShunBl13} that provides a benchmark suite of over 20 non-trivial graph problems.
GBBS uses a shared-memory approach to parallel graph processing in
which the entire graph is stored in the main memory of a single
multicore machine.
Graphs in GBBS are assumed to be stored in the compressed sparse row (CSR)
format described earlier. The underlying neighbors stored can be stored either
uncompressed, or using a compressed format.

\paragraph{Vertex datatypes and primitives}
The vertex datatype interface
provides functional primitives over vertex neighborhoods, such as
map, reduce, scan, count (a special case
of reduce where the map function is a boolean function), as well as
primitives to extract a subset of the neighborhood satisfying a
predicate (filter), among other primitives.

\paragraph{VertexSubsets}
GBBS uses the vertexSubset datatype from Ligra, which represents a subset of
vertices in the graph. A subset can either be \emph{sparse}
(represented as a collection of vertex IDs) or \emph{dense}
(represented as a boolean array or bit-vector of length $n$, the
number of vertices in the graph).

\paragraph{EdgeMap}
\edgemap is a basic
graph processing primitive useful for performing graph traversal.
The \edgemap primitive takes as input a frontier, or vertexSubset. It then applies a user-defined function to generate a new
frontier consisting of neighbors of the input frontier. For example,
in a breadth-first search, the user-defined primitive emits a neighbor
in the output frontier if it has not yet been visited.
GBBS includes several generalizations of \edgemap that aggregate the results
of the \edgemap at the source vertex, as well as generalizations that
return a subset of the neighbors of the input vertexSubset.


\section{BYO API design and implementation}\label{sec:byo-api}
\begin{figure}
  \centering
  \includegraphics[width=.8\linewidth]{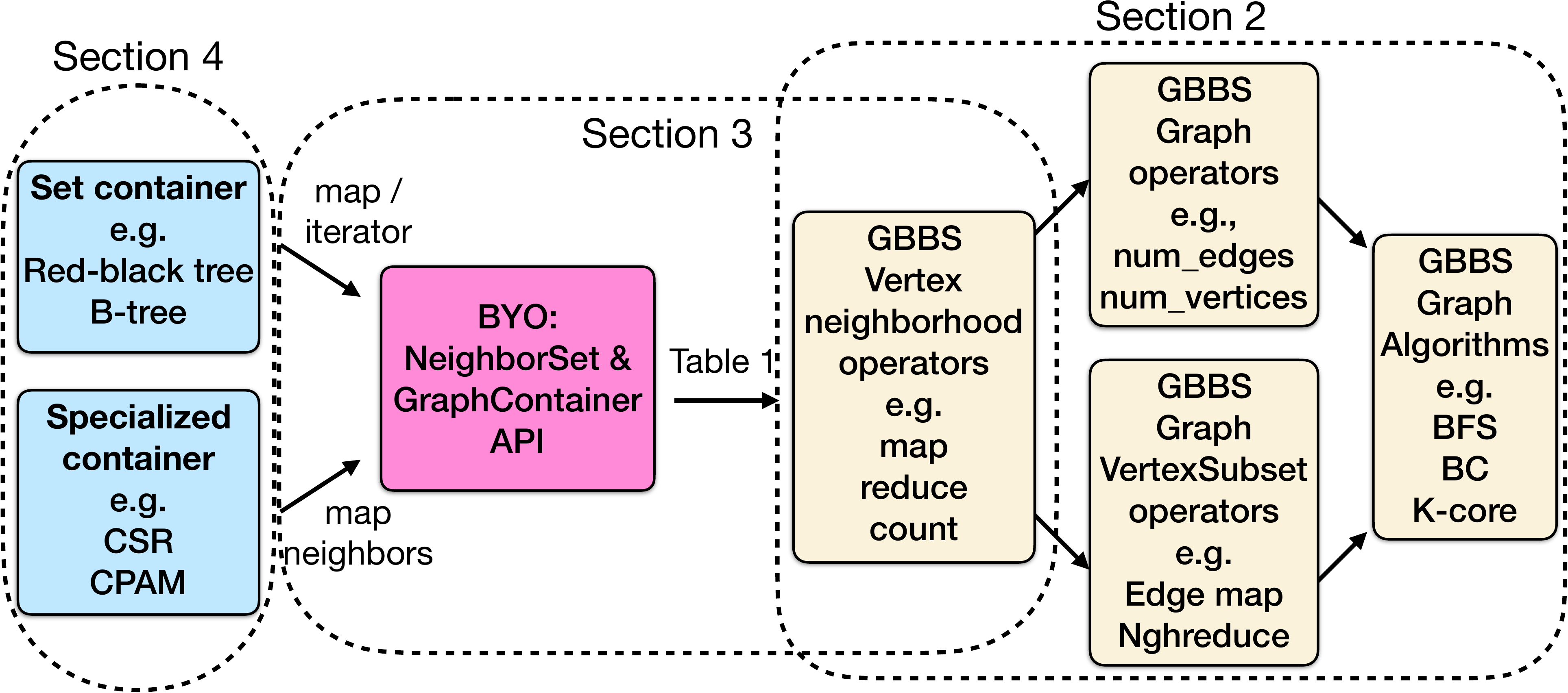}
  \caption{Relationship between \apinameshort framework, graph containers, and graph algorithms (via GBBS).}
  \label{fig:api-design}
\end{figure}

The goal of \apinameshort is to make it as easy as possible for a
graph-container developer to use any data structure in a high-performance and
general graph programming framework. This section details the ``set'' and
``graph container'' APIs that \apinameshort exposes to connect with arbitrary
graph data structures.  It also describes the changes that we made to the GBBS
implementation to be agnostic to the underlying data structure. Finally, we will
discuss framework-level optimizations that have appeared in various places
throughout the literature that we have collected in \apinameshort. For
simplicity, we will describe all of the API components in terms of unweighted
graphs.

\apinameshort provides a translation layer between graph containers and
GBBS.
~\figref{api-design} illustrates the relationship between data structures,
\apinameshort, and GBBS components. Specifically, \apinameshort translates
between the data structure API and the read-only GBBS neighborhood operators
such as map, reduce, and scan (\secref{gbbs-api}). This paper focuses on these
operators to strike a balance between simplicity and expressiveness.

\subsection{\setapi{}}\label{sec:set-api}

As described in~\secref{prelim}, a graph can be represented as a sequence of
sets where each set contains the edges incident to a single vertex, that is a
neighbor set.  Many graph representations, such as the adjacency list in
Stinger~\cite{EdgigerMcRi12}, the tree of trees in Aspen~\cite{DhulipalaBlSh19}
and CPAM~\cite{DhulipalaBlGu22}, directly implement this two-level structure.

We now describe the \defn{\setapi{}}, a high-level description of the necessary
functionality for \emph{a single} vertex neighbor set. The \setapi{} enables
easy parallelization over the vertex set, since all of the neighbor sets are
independent.

\begin{figure}
  \centering \includegraphics[width=.8\linewidth]{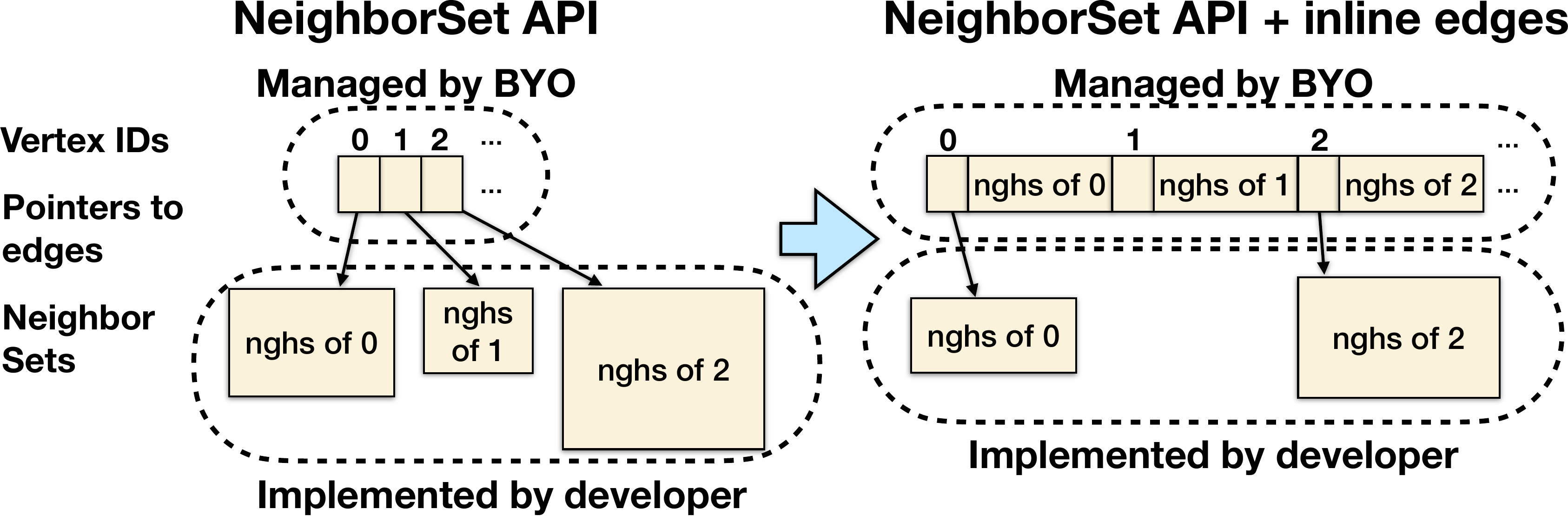}
  \caption{Data structure and inline optimization using the set
    API.}
  \label{fig:set-api-design}
\end{figure}

~\figref{set-api-design} illustrates the relationship between the vertex level
(maintained by \apinameshort) and the set data structures (implemented by the
developer). \apinameshort abstracts away the details of choosing a data
structure for both the vertex sequence and neighbor sets and enables the user to
just implement the neighbor set.  Currently, \apinameshort implements the vertex
sequence as an \texttt{std::vector} for simplicity, but could theoretically use
any set data structure.

We find that the minimal API necessary for a neighbor set data structure to
implement the GBBS operators which do not change the neighbor sets is to expose
a \texttt{size} function and an \texttt{iterator} which supports sequentially
iterating through the elements one by one, i.e., a \defn{forward} iterator.
These are two basic functionalities are naturally expected from set
implementations. For example, the \texttt{C++} standard template library
(STL)~\cite{MusserDeSa01, Josuttis12}, a widely-used standard library of basic
utilities, includes both of these (among others).

\paragraph{Required functionality for algorithms}
\begin{itemize}
\item \texttt{iterator} or \texttt{map(f)}: Apply the function \texttt{f} to all
  elements in the set. As mentioned in~\secref{intro}, an iterator can be used
  to implement \texttt{map} by simply iterating through all elements in the set
  and applying the function \texttt{f}.
\item \texttt{size()}: Return the number of elements in the set.
\end{itemize}

\paragraph{Optional functionality for algorithms}

\begin{itemize}
\item \texttt{map\_early\_exit(f)} (if no iterator): Apply the function
  \texttt{f} to the elements in the set until \texttt{f} returns true. We can
  implement it with an iterator by exiting the iteration
  when \texttt{f} returns true.
\item \texttt{parallel\_map(f)}: Apply the function \texttt{f} in parallel to
  all elements in the set.
\item \texttt{parallel\_map\_early\_exit(f)}: Apply the function \texttt{f} in
  parallel to the elements in the set until \texttt{f} returns true. Because the
  iterations are running in parallel, other threads may continue even after one
  returns true.
\end{itemize}

\paragraph{Required functionality for updates}
In addition to providing read access to the graph, dynamic-graph representations
must also support updates (inserts / deletes). Therefore, \apinameshort requires
the following standard API if the data structure supports dynamicity:

\begin{itemize}
\item \texttt{insert(e)} : Insert a single element \texttt{e} into the set.
\item \texttt{delete(e)}: Delete a single element \texttt{e} from the set.
\end{itemize}

\paragraph{Optional functionality for updates}
Some data structures (e.g., Aspen~\cite{DhulipalaBlSh19} and
CPAM~\cite{DhulipalaBlGu22}) may natively support batch updates at the set
level. That is, in addition to the parallelization over the vertices, the data
structure itself may support batch updates for work sharing and potentially
additional parallelism. Therefore, \apinameshort also provides an interface for
batch insertions/deletions as part of the \setapi{}:

\begin{itemize}
\item \texttt{insert\_batch(batch)}: Insert a batch of elements
  into the set.
\item \texttt{delete\_batch(batch)}: Delete a batch of elements
  from the set.
\end{itemize}

\paragraph{Translating graph batch-update API into \setapi{}}
\rev{
Modern dynamic-graph systems support inserting and deleting a \defn{batch}
(i.e., a set) of edges rather than one at a time. Different data structures
require different input forms for efficient batch updates to be applied.
\apinameshort supports three different forms for batches.  The first is to simply globally sort the batch.  This is good for data structures that perform global merges.  The second is to semi sort~\cite{van1991handbook, gu2015top, dong2023high}, which groups equal
keys (sources) together but does not necessarily globally order the keys across
the whole list, to partition the batch into edges destined for different
vertices the batch by source. Just semi sorting is sufficient for data structures that use the \setapi{}, but do not derive any benefit from sorting such as hash tables.  The third is to first semi sort by source, and then group and integer sort each individual set of edges.  This is good for ordered set containers.
}


\paragraph{Advantages of the \setapi{}}
The \setapi{} is designed to make it as easy as possible for a data-structure
developer to integrate their container with \apinameshort, as long as they
implement the basic contract specified in the STL container API.  Notably, if a
developer wants to integrate a set library that implements \texttt{size} and
\texttt{iterator} functionality with \apinameshort, they \emph{do not need to
  write any additional code}. To improve ease of use, we implemented
\apinameshort to automatically translate from the STL container API to the
\apinameshort API. That is, integrating a data structure that implements the STL
container API just requires importing it at the top of the test driver and
specifying its type as the graph container under test.

Furthermore, we incorporate the \defn{inline optimization} from
Terrace~\cite{PandeyWhXu21} into the vertex set in \apinameshort to benefit
arbitrary data structures and enable faster systems overall. This optimization
stores a few (about 10) edges inline in the vertex level next to the pointer to
the neighbor set for each vertex. The goal is to avoid indirections for
low-degree vertices. The idea was originally introduced in Terrace but can be
generally applied to any graph container with the sequence of sets structure.
~\figref{set-api-design} illustrates the inline optimization in an arbitrary
sequence of sets graph representation.  On average, we find that the inline
optimization speeds up set containers by \inlineavgspeedup on average, which we
will detail in~\secref{eval}.

Another benefit of the \setapi{} is flexibility in the choice of outer set data
structure with a fixed inner set type. For example, \apinameshort can store
directed graphs with two dense outer vectors (one for incoming and one for
outgoing neighbors). Similarly, \apinameshort could store the outer set sparsely
for a static or dynamic form of Doubly-Compressed Sparse Rows
(DCSR)~\cite{buluc2008representation}.

\subsection{\containerapi{}}\label{sec:graph-api}
Next, we introduce the \defn{\containerapi{}} to connect \apinameshort to
graph data structures that do not represent the neighbor sets as separate
independent data structures. For example, the classical Compressed Sparse Row
(CSR)~\cite{TinneyWa67} representation stores all of the neighbor sets
contiguously in one array. Furthermore, some optimized dynamic-graph
containers like Terrace~\cite{PandeyWhXu21} and SSTGraph~\cite{WheatmanBu21}
collocate some neighbor sets for locality. These graph data structures
internally manage both the vertex and neighbor sets.

We find that the minimal API necessary for a data structure to support a diverse
set of graph algorithms via \apinameshort is just \texttt{map\_neighbors} and
\texttt{num\_vertices}.

\paragraph{Required functionality for algorithms}
\begin{itemize}
\item \texttt{map\_neighbors(i, f)}: Apply the function \texttt{f} to all
  neighbors of vertex \texttt{i}.
\item \texttt{num\_vertices()}: Return the number of vertices in the graph.
\end{itemize}

\paragraph{Optional functionality for algorithms}

\begin{itemize}
\item \texttt{num\_edges()}: Return the number of edges in the graph.
\item \texttt{degree(i)} : Return the degree of vertex \texttt{i}.
\item \texttt{map\_neighbors\_early\_exit(i, f)}: Apply the function \texttt{f}
  to the neighbors of vertex \texttt{i} until \texttt{f} returns true.
\item \texttt{parallel\_map\_neighbors(i, f)}: Apply the function \texttt{f} in parallel to
  all neighbors of vertex \texttt{i}.
\item \texttt{parallel\_map\_neighbors\_early\_exit(i, f)}: Apply the function
  \texttt{f} in parallel to the neighbors of vertex \texttt{i} until \texttt{f}
  returns true. Because the iterations are running in parallel, some threads
  may continue even after one returns true.
\end{itemize}

Many of these functions such as \numvertices, \numedges, and \degree are
commonly expected from any graph container as a way to query the graph
structure. The \texttt{map\_neighbors} functionality is also commonly
implemented in graph containers to support graph algorithms. The optimized
variants of map (early exit and parallel) can be more difficult to implement
than serial map. They have been studied in the literature~\cite{ShunGhBl15,
  DhulipalaBlSh19, BeamerAsPa12} and have been reported to help on some
algorithms and graphs. We perform a comprehensive study of the performance
benefits of the individual API components in~\secref{eval}.

\paragraph{Required functionality for updates}
The \containerapi{} directly translates the batch-update API at the
\apinameshort level to the underlying container. Since the batch given to
\apinameshort may not be sorted, \apinameshort sorts it because many
batch-update algorithms require sorted batches~\cite{PandeyWhXu21,
  DhulipalaBlSh19, DhulipalaBlGu22, WheatmanBuBu23, WheatmanBu21}. Data
structures using the \containerapi in \apinameshort must implement the following
functions:

\begin{itemize}
\item \texttt{insert\_sorted\_batch(batch)}: Insert a sorted batch of edges into
  the graph.
\item \texttt{delete\_sorted\_batch(batch)}: Delete a sorted batch of edges from
  the graph.
\end{itemize}

\paragraph{Advantages of the \containerapi{}}
The \containerapi{} enables cross-set optimizations that cannot be expressed in
the \setapi{} at the cost of programming effort. For example, in the classical
CSR, the edges are stored contiguously in one array for locality rather than in
separate per-vertex arrays, which is not easily captured by the set of sets
abstraction. Another example is the hierarchical structure in
Terrace~\cite{PandeyWhXu21}, which stores some neighbor sets contiguously in a
dynamic array-like data structure. Additionally, SSTGraph~\cite{WheatmanBu21}
shares some metadata between the different neighbor sets for space savings,
which cannot be captured with the independent sets abstraction. However, the
\containerapi cannot access the general inline optimization supported by the
\setapi.

\subsection{Maintaining graph metadata in \apinameshort}\label{sec:byo-metadata}
Furthermore, \apinameshort reduces the burden on the programmer by internally
maintaining information about graph structure at the framework level when it is
not done at the container level. Specifically, it stores the the degree of each
vertex in the vertex set as well as the total number of edges if needed. If a
set container does not implement \texttt{size} or a graph container does not
implement \texttt{degree} and/or \texttt{num\_edges}, \apinameshort defaults to
its internal metadata.

\subsection{Connecting \apinameshort to GBBS}

\begin{table}
\scriptsize
  \centering
  \begin{tabular}{@{}lr@{}}\toprule\begin{tabular}{@{}l@{}}\textit{GBBS Vertex} \\
                                     \textit{Operator}  \end{tabular}         & \textit{B.Y.O. Lambda} \\
    \midrule
    Map & Pass through provided function \\[3pt]
    Reduce & auto value = identity \\
                                                                              & map([\&](auto ...args) \{ value.combine(f(args...)) \})\\[3pt]
    Count & int cnt = 0 \\
                                                                              &  map([\&](auto ...args) \{ cnt += f(args...) \})\\[3pt]
    Degree & int cnt = 0  \\
                                                                              & map([\&](auto ...args) \{ cnt += 1) \})\\[3pt]
    getNeighbors & Set ngh = \{\}  \\
                                                                              & map([\&](auto ...args) \{ ngh.add(args) \})\\[3pt]
    \new{ Filter} & \new{Set ngh = \{\} }  \\
                                                                              & \new{map([\&](auto ...args) \{ if (pred(args) ngh.add(args) \})}\\[3pt]
    \bottomrule
  \end{tabular}
  \caption{GBBS primitives implemented using just the map
    primitive.
  }
  \label{tab:gbbs-byo-translation}
\end{table}

\apinameshort simplifies the list of original read-only GBBS neighborhood
operators such as map, reduce, count, etc. by implementing several of them with
\texttt{map}. The original GBBS specification required the data-structure
developer to implement several neighborhood operators. In contrast,
\apinameshort requires them to implement {\bf only
  one}. ~\tabref{gbbs-byo-translation} demonstrates how to implement the
original GBBS neighborhood operators using different \texttt{map} lambdas.

In addition to providing the translation layer from the GBBS vertex neighborhood
operators, we also needed to modify the implementation of some vertexSubset
operators in GBBS because they assume that the underlying graph is stored in CSR
format. This is not inherent in the high-level GBBS specification, but was a
prevalent assumption in the codebase. Specifically, several EdgeMap functions
assumed that they could directly perform array access into the container to
access relevant parts of the graph, which does not hold for arbitrary data
structures. 

\paragraph{VertexSubset optimizations in \apinameshort}
We also include several optimizations to the higher-level vertexSubset
abstraction in \apinameshort that can benefit all systems, since they are
independent of the container.  Specifically, we converted the boolean array that
the vertexSubset uses in dense mode to a {\em bitarray}, which has been shown in
prior work~\cite{WheatmanBu21} to improve overall algorithm performance by about
$1.05\times$. We confirm these results with our own experiments and find that on
average, \apinameshort with a boolean array is $1.04\times$ slower than
\apinameshort with a bit array.  We also removed some unnecessary work (e.g.,
copies and sorts) at the vertexSubset level.

\paragraph{GBBS algorithms supported in \apinameshort} \mymarginpar{R4D2}
\rev{In addition to algorithms that only require read-only neighborhood
  operators (e.g., map, reduce, degree, etc.), \apinameshort can also support
  whole-graph algorithms that at first seem to require vertex-vertex operators such as
  intersection by first applying an out-of-place \emph{filter} operation (see
  ~\tabref{gbbs-byo-translation}) to convert the data structure into a more
  amenable format for algorithms.
  For example, efficient implementations of
  triangle counting (TC), a classical example of an algorithm based on
  vertex-vertex intersections, first perform a filter to reduce the number of
  edges and eliminate enumerating duplicate triangles~\cite{7113280}.
  Applying this filtering optimization is the standard technique for running
  other whole-graph algorithms involving vertex-vertex operations, e.g., k-truss,
  butterfly counting, structural
  similarities, etc~\cite{che2020accelerating, fang2019efficient, kabir2017parallel, sanei2018butterfly, shi2022parallel, xu2007scan}.

  Notably, this filtering operation is not an algorithmic change made in \apinameshort. The triangle
  counting (TC) implementation in base GBBS first performs a filter before
  intersection, even when the graph is stored in CSR.  To demonstrate
  \apinameshort's functionality, we used filter to implement TC on top of a
  variety of dynamic containers. We found that they all support TC in similar
  time because the only variation is the time spent doing the filter, which is relatively
  inexpensive compared to the time performing intersections on the filtered graph.

  }


\section{Data structures evaluated}\label{sec:datastr}
This section details the many graph containers that we evaluate using
\apinameshort in this paper. We include general-purpose off-the-shelf data
structure libraries that were not designed or optimized for graph processing as
easy-to-use baselines. We also include state-of-the-art special-purpose graph
containers from the literature to demonstrate the advantage of further
development and optimization.


\subsection{Off-the-shelf data structures}\label{sec:general-purpose}
We evaluate off-the-shelf data structures both from the \texttt{C++} standard
template library (STL)~\cite{MusserDeSa01, Josuttis12}, and from
Abseil~\cite{absl}, a collection of open-source data structure implementations
in \texttt{C++} designed to augment the STL. These containers
serve as baselines to demonstrate how much performance is left on the table with
simple easy-to-use data structures. The details about the included structures
are as follows:

\begin{itemize}
\item \stdset~\cite{stdset}: A standard container library that comes
  with any \texttt{C++} distribution. It maintains a sorted set of unique
  elements and is usually implemented with a red-black tree~\cite{CLRS}.

\item \stdunorderedset~\cite{stdunorderedset}: A standard container
  library introduced in \texttt{C++11}. The elements are stored unsorted in a
  hash table~\cite{CLRS}.

\item \abslbtree~\cite{absl}: An ordered set data structure that
  generally conforms to the STL container API.  It is implemented as a
  B-tree~\cite{BayerMc72}, a classical cache-friendly tree data structure.

\item \abslflathash~\cite{absl}: An unordered set data structure that generally
  conforms to the STL container API. It is implemented using a hash table.

\item \stdvector~\cite{stdvector}: A basic array implementation that
  comes with any \texttt{C++} distribution. It stores elements continguously in
  memory.
\end{itemize}

\paragraph{Integration with \apinameshort}
These general-purpose data structures are all sets and therefore use the \setapi
described in~\secref{set-api}. Using the common STL container API functions of
\texttt{size} (returns the number of elements in the container) and
\texttt{iterator} (enables access to the elements in the container), the
general-purpose data structures naturally support the \texttt{num\_edges},
\texttt{degree}, \texttt{map}, and \\ \texttt{map\_with\_early\_exit}
functionality through translation via \apinameshort.

Since these set data structures implement the \setapi, they do not need any
additional code to integrate with \apinameshort, as described
in~\secref{set-api}.  However, vector is still implemented in full manually to
give users of the system an example of each function being implemented and to
include the parallel mapping functions which can not be generated
automatically.

\subsection{Optimized data structures}\label{sec:special-purpose}

We also evaluate optimized general set data structures that have previously been
used for dynamic graphs as well as special-purpose graph containers designed
specifically for graphs. These containers demonstrate what kind of performance
is achievable with tailor-made data structures. Some of these optimized data
structures have both uncompressed and compressed versions (\secref{prelim}),
which we will denote with $^\dagger$. The details are as follows:

\begin{itemize}

\item Compressed Sparse Row$^\dagger$~\cite{TinneyWa67} (CSR): A classical
  \emph{static} representation for graphs.
\item Terrace~\cite{PandeyWhXu21}: A dynamic-graph container optimized for
  skewed graphs. It uses a hierarchical structure built on arrays, a Packed
  Memory Array (PMA) for graphs~\cite{WheatmanXu21, WheatmanXu18}, and
  B-trees~\cite{BayerMc72} to organize the vertices by degree.
\item SSTGraph~\cite{WheatmanBu21}: A dynamic-graph container built on a shallow
  hierarchy of sorted PMAs~\cite{itai1981sparse, BendDeFa00}.
\item PMA$^\dagger$~\cite{WheatmanBuBu23}: A cache-oblivious updateable array.
\item Dynamic Hashed Blocks (DHB)~\cite{VanPrWi22}: A dynamic-graph container
  based on block-based hashing.
\item Aspen$^\dagger$~\cite{DhulipalaBlSh19}: A randomized blocked tree.
\item CPAM$^\dagger$~\cite{DhulipalaBlGu22}: A deterministic blocked tree.
\end{itemize}

\paragraph{Integration with \apinameshort}
We integrate CSR~\cite{TinneyWa67}, Terrace~\cite{PandeyWhXu21},
SSTGraph~\cite{WheatmanBu21}, and DHB~\cite{VanPrWi22} with the \containerapi in
\apinameshort. All of these data structures already natively include the
\texttt{degree} functionality, and most include \texttt{num\_edges}. For the
systems that implemented the Ligra interface (Terrace~\cite{PandeyWhXu21},
SSTGraph~\cite{WheatmanBu21}, and PMA~\cite{WheatmanBuBu23}), we adapted the
map-based functionality from their original integration with Ligra to integrate
with \apinameshort. In DHB~\cite{VanPrWi22}, we used the provided iterator to
implement the \texttt{map\_neighbors} and\\ \texttt{map\_neighbors\_early\_exit}
functionality.

Additionally, we incorporate Aspen~\cite{DhulipalaBlSh19} and
CPAM~\cite{DhulipalaBlGu22} with the \setapi in \apinameshort. These tree-based
containers natively implement \texttt{size} and all of the map variants. We
incorporate the PMA~\cite{WheatmanBuBu23} into both the NeighborSet and
\containerapi in \apinameshort since it was originally presented as a single PMA
for the entire graph.


\section{Experimental evaluation}\label{sec:eval}

We employ \apinameshort to evaluate \numdatastructures graph containers (and
variants), \numalgorithms graph algorithms, and \numgraphs graph datasets.


We first summarize the high-level takeaways from our large-scale
evaluation. Next, we study the performance effects of the different
functionalities in the \apinameshort API described in~\secref{byo-api}, which
helps explain the impact of missing functions in later evaluations. We then
evaluate \apinameshort compared to other state-of-the-art graph-algorithm
frameworks to ensure that \apinameshort's extra generality does not come at the
cost of performance. Finally, we perform a comprehensive evaluation of
\numdatastructures graph data structures on a suite of \numalgorithms graph
algorithms. We also evaluate the dynamic structures on their update throughput.
We publicize the raw data for all experiments as well as the code in the Github
repo\footnote{\url{https://github.com/wheatman/BYO}}.  \new{We evaluate all
  systems in an unweighted, undirected mode to enable the widest compatibility.}



\subsection{Summary}\label{sec:summary}

First, we evaluate different configurations of the \containerapi
(~\secref{graph-api}) in \apinameshort with CSR as the underlying graph
representation to determine the importance of the different functions (e.g., the
types of maps). On average, we find that the \defn{minimal efficient} API
configuration implements the required functionality, \degree, and
\numedges. This configuration is only \minefficientslowdown slower compared to
the \defn{full configuration}, or the \apinameshort configuration with all of
the (required and optional) functionality listed in~\secref{graph-api}. However,
in the worst case, the minimal efficient configuration incurs up to $3.1\times$
slowdown over the full API. Adding the more advanced maps (parallel map and
early exit) mitigates the worst case, which is important for difficult problem
instances e.g., graphs with high degree.  The full details are
in~\secref{byo-micros}.

Next, we evaluate \apinameshort compared to other state-of-the-art
graph-algorithms frameworks in~\secref{frameworks-eval} and find that
\apinameshort is competitive with GraphBLAS, Ligra, and GBBS, which are
state-of-the-art high-performance frameworks for graph algorithms. Specifically,
\apinameshort achieves between $1.06-4.44\times$ speedup on average compared to
other frameworks.
These results indicate that \apinameshort is a good candidate for integrating
with various graph containers because the resulting systems are both expressive
and achieve high performance.  We also compare several of the original systems
that introduced dynamic-graph containers (e.g., PMA, CPAM) with their original
frameworks to their implementations using BYO and find that BYO's implementation
is faster.

Finally, we perform a comprehensive study of dynamic-graph containers on both
graph algorithm and batch-insert performance
in~\secreftwo{algs-eval}{batch-eval}.

In terms of graph algorithm performance, our findings show that graph data
structures are very similar on average, but that developing specialized graph
data structures is worthwhile because additional optimization effort can improve
holistic performance on more challenging instances, e.g., high-degree
graphs. All of the data structures tested besides the unoptimized \stdset and
\stdunorderedset incur most about \avgmaxslowdown slowdown compared to CSR when
averaging across all algorithms and graphs. Furthermore, the best specialized
container (CPAM with inline) is only about $1.1\times$ faster than the best
off-the-shelf data structure (\abslbtree with inline) on average. However, the
worst-case slowdown for the \abslbtree is $2.6\times$, while CPAM achieved a
maximum slowdown of $1.9\times$ over CSR. These results suggest that specialized
data structures can improve upon off-the-shelf data structures on more difficult
problem settings.

\apinameshort cuts through combinatorial explosion in terms of programming
effort to enable large-scale comparisons of graph containers on a diverse suite
of algorithms to provide a complete view of how fast a graph container can
support algorithms in a variety of cases.

In terms of
batch inserts, we find that off-the-shelf structures exhibit a folklore
query-update tradeoff: the Abseil B-tree, which is best off-the-shelf structure
for algorithms, \rev{experienced around a $3\times$ slowdown on larger batch inserts}
compared to the Abseil flat hash set. However, the hash set was worse on
algorithms compared to the B-tree. However, specialized containers can overcome
the query-update tradeoff on the largest batches: the single PMA is better on
algorithms on average compared to the B-tree as well as on the largest batch
size.

\begin{table}[t]
  \centering
  \scriptsize
  \begin{tabular}{@{}llr@{}}
    \toprule
    \textit{Category}          &   \textit{Problem}   \\
    \midrule
   \multirow{3}{*}{ \begin{tabular}{@{}l@{}}Shortest-path \\
                          problems\end{tabular}}  &    Breadth-First Search (BFS) & \\
   &  Single-Source Betweenness Centrality (BC)  & \\
                               & $O(k)$-Spanner (Spanner)  & \\
    \hline
 \multirow{2}{*}{Connectivity}  & Low-Diameter Decomposition (LDD) \\
                               & Connectivity (CC) \\
     \hline
  \multirow{2}{*}{Substructure} & Approximate Densest Subgraph (ADS) \\
                               & $k$-core \\
     \hline
   \multirow{2}{*}{Covering} &  Graph Coloring (Coloring) \\
                               &   Maximal Independent Set (MIS) \\
     \hline
 Eigenvector &  PageRank (PR) \\
    \bottomrule
  \end{tabular}
  \caption{Graph problems supported in \apinameshort and their categories~\cite{DhulBlSh21}.}
  \label{tab:algs-list}
\end{table}

\subsection{Experimental setup}
\paragraph{Algorithms evaluated}
\tabref{algs-list} lists the \numalgorithms graph problems that \apinameshort provides
parallel algorithms for based on the data-structure API and GBBS abstractions.
\apinameshort does not change the algorithm
implementations from GBBS (just the translation from the data structure to the lower-level primitives). Therefore, \apinameshort inherits the strong theoretical bounds on algorithm work and depth (and therefore
parallelism) from GBBS. These algorithms cover a wide range of problems, including shortest-path,
connectivity, substructure, covering, and eigenvector problems. We refer
the interested reader to the GBBS paper for full details on the algorithms and
their implementations~\cite{DhulBlSh21}.

\paragraph{Systems setup}
We ran all experiments on an Intel\textregistered Xeon\textregistered Gold 6338
CPU @ 2.00GHz dual socket machine
with 64 physical cores (128 hyperthreads) and 1024 GiB of
main memory running across 16 channels at 3200MT/s. The machine has 5 MiB of L1 cache, 80 MiB of L2 cache, and
96 MiB of L3 cache.

To validate the choice of the \apinameshort API, and show that it achieves similar end-to-end running times as existing systems, we first compare \apinameshort with Ligra, GBBS, GraphBLAS, and GAP on the 4 algorithms common to all of the systems: BFS, BC, CC, and PR (\tabref{algs-list}). Ligra/GBBS/GraphBLAS are state-of-the-art graph-algorithm frameworks, while GAP is a suite of direct algorithm implementations.

We compiled all codes besides Ligra~\cite{ShunBl13} with \texttt{g++ 11.4.0}.
GraphBLAS and GAP~\cite{BeamAsPa15} are parallelized natively with
OpenMP~\cite{DagumMe98}. Since Ligra~\cite{ShunBl13} was designed and tested
with Cilk~\cite{BlumofeJoKu95} but Cilk is no longer supported in \texttt{g++},
we compiled Ligra using \texttt{clang++ 14.0.6} with OpenCilk
2.0~\cite{SchardlLe23}, the modern iteration of Cilk.  We ran GBBS
with its default custom parallelization framework and
scheduler. \apinameshort uses the same custom parallel scheduler since it
uses GBBS as a starting point for its implementation.

To test the GraphBLAS programming framework~\cite{KepnerAaBa16, BulucMaMc17,
  Davis19, Davis23}, we include LAGraph~\cite{lagraphgit, MattsonDaKu19}, a
collection of algorithms implemented using GraphBLAS.

We kept the number of trials unchanged from each system's distribution and took
the average over all trials. By default, Ligra/GBBS/\apinameshort perform 3
trials (with one extra warmup trial) per experiment, GAP performs 16, and GraphBLAS performs 64.

Furthermore, we evaluate the systems resulting from integrating \apinameshort with existing containers compared to their original systems.
Specifically, we ran PMA, SSTGraph and CPAM with their original driver code, compiled with g++-11, which use their own implementation of the Ligra API to run algorithms. The original systems use the same parallelization framework as \apinameshort. We ran the PMA variants and SSTGraph on BFS, BC, PR, and CC, and CPAM on BFS and BC because those were the algorithm implementations provided in the original codebases that intersected with the ones provided via \apinameshort. The algorithm implementations vary slightly between \apinameshort and the existing systems, but the high-level abstractions are the same.

We ran all graph containers in unweighted mode (and by extension, all algorithms) for simplicity.

\begin{table}
\scriptsize
  \centering
  \begin{tabular}{@{}llrr@{}}\toprule
    \textit{Graph}         & \textit{Vertices} & \textit{Edges} & \textit{Avg. Degree}  \\
    \midrule
    Road (\roadshort)           &23,947,347&57,708,624&2\\
    LiveJournal (\ljshort) & 4,847,571&85,702,474&18\\
    Com-Orkut (\orkutshort) &3,072,627&234,370,166&76\\
    rMAT (\rmatshort) & 8,388,608 & 563,816,288 & 67 \\
    Erd\H{o}s-R\'enyi (ER) & 10,000,000&1,000,009,380&100\\
    Protein (PR) & 8,745,543 & 1,309,240,502 & 150 \\
    Twitter (TW)           &61,578,415&2,405,026,092&39\\
    papers100M (PA) & 111,059,956 & 3,228,124,712& 29 \\
    Friendster (FS)        & 124,836,180&3,612,134,270&29\\
    Kron (KR)           &134,217,728&4,223,264,644&31\\
    \bottomrule
  \end{tabular}
  \caption{Sizes of (symmetrized) graphs used (ordered by size).}
  \label{tab:graph-sizes}
\end{table}

\paragraph{Datasets}
\tabref{graph-sizes} lists the \numgraphs graphs used in the evaluation and
their sizes. All of the graphs included in the evaluation are undirected and
unweighted for simplicity. We started with a selection of graphs from the GAP
benchmark suite~\cite{BeamAsPa15}, a widely-used graph-evaluation
specification. These include the the Road (RD)~\cite{road} network, the
\defn{Twitter} (TW)~\cite{BeamAsPa15} graph, and the \defn{Kron}
graph~\cite{LeskovekChKl05} with the same parameters as Graph 500 (A=0.57,
B=C=0.19, D=0.05)~\cite{graph500}.  Much like in GAP, we include a uniform
random graph \defn{Erd\H{o}s-R\'enyi} (ER) graph~\cite{erdos59a} generated with
$n = 10^7$ and $p = 5\cdot 10^{-6}$. We also generated an \defn{rMAT} (RM) graph
by sampling edges from an rMAT generator~\cite{ChakZhFa04} with
$a = 0.5; b = c = 0.1; d = 0.3$ to match the distribution commonly used in
evaluating graph containers~\cite{DhulipalaBlSh19, PandeyWhXu21,
  DhulipalaBlGu22, WheatmanBuBu23}. Additionally, we include several other
social network graphs: the \defn{LiveJournal} (LJ)~\cite{LJ}, \defn{Community
  Orkut} (CO)~\cite{orkut}, and \defn{Friendster} (FS) graphs from the SNAP
dataset~\cite{snapnets}. For coverage from other application domains, we include
the \defn{papers100M} (PA) dataset from the Open Graph
Benchmark~\cite{HuFeZi20}. Finally, we also use the \defn{Protein} (PR) network
graph, an induced subgraph available in the data repository of
the HipMCL~\cite{AzadPaOu18} algorithm. Unlike social network graphs, the protein network graph is
not heavily skewed in terms of degree distribution. These inputs represent a wide range of inputs both in
skewness and in size.

\begin{table}
\scriptsize
\begin{tabular}{@{}lrrr@{}}
\toprule
\textit{API configuration} & \multicolumn{3}{c}{\textit{Slowdown over full API}} \\
                            &\textit{Average}  & \textit{95\%} & \textit{Max}  \\
\midrule
Min (just \mapneighbors and \numvertices) & 10.69 & 231 & 1379  \\
Min + \degree & 1.43 & 4.1 & 22.8 \\
Min efficient (Min + \degree + \numedges) &\cellcolor{safe-green!25} 1.16 & 2.5 & 3.1 \\
Full minus \numedges & 1.31 & 2.78 & 22.9 \\
Full minus \degree & 2.18 & 7.4 & 14.5  \\
No early exit (Full minus both map early exit) & \cellcolor{safe-green!25} 1.12 & 2.5 & 3.1 \\
No parallel map (Full minus both parallel map) & \cellcolor{safe-green!25} 1.01 & 1.3 & 1.9 \\
Full minus \texttt{parallel\_map\_neighbors\_early\_exit} & \cellcolor{safe-green!25} 0.98 & \cellcolor{safe-green!25}1.03 & \cellcolor{safe-green!25}1.1 \\
Full (All required and optional functionality) & \cellcolor{safe-green!25} \cellcolor{safe-green!25}1.00 &\cellcolor{safe-green!25}1.00 & \cellcolor{safe-green!25}1.00 \\
\bottomrule
\end{tabular}
\caption{The average performance of different \containerapi configurations with CSR as the underlying container. ``Min'' refers \apinameshort with just the required functionality and ``Full'' refers to \apinameshort with both required and optional functions described in Section~\ref{sec:graph-api}.  The remaining configurations are described with either what they add to min, or what they remove from full.  The 95\% and max columns show the 95th percentile and maximum slowdown over the full API across all algorithms and all graphs. Configurations that achieve within $1.25\times$ slowdown over the full API are shaded.}
\label{tab:api-components-avg}
\end{table}

\subsection{API Microbenchmarks}\label{sec:byo-micros}
We start by performing a study to understand what
are the important ways in which the graph processing system needs to interact
with the underlying graph.
We perform a set of experiments where we keep the graph data
structure consistent (CSR) and only vary the API we use to run the algorithms.
~\tabref{api-components-avg} reports the average slowdown relative to the full
API (all functions implemented) of the different API configurations (missing
some functions).

The \defn{minimal efficient} API configuration with only \texttt{num\_edges} and
\texttt{degree} on top of the required functionality (\texttt{map} and
\numvertices), \apinameshort incurs $1.16\times$ slowdown on average compared to
the full API with all map variants. Furthermore,
In terms of
algorithms, 
the minimal efficient API incurs over $1.2\times$ slowdown (on average across
graphs) on BFS, Spanner, and LDD, which can be explained due to the lack of
early exit, as these algorithms can all benefit from direction-optimization and
early termination during edgeMap~\cite{BeamerAsPa12}.  These results suggest
that a data-structure developer that implements only a few basic functions such
as \numedges and \degree can achieve close to the best-possible performance from
\apinameshort on a majority of cases.
The \numedges and \degree functions are necessary for performance because they
are used frequently in the \edgemap/\vertexsubset graph-algorithm abstraction
from Ligra/GBBS to determine the cutoff between sparse and dense mode for the
\vertexsubset~\cite{ShunBl13}.
%

Furthermore,~\tabref{api-components-avg} shows that implementing the more
advanced maps does not significantly help: on average, omitting map with early
exit incurs $1.12\times$ slowdown and omitting parallel map incurs only
$1.01\times$ slowdown. However, there are specific algorithms and graphs on
which these advanced maps are critical for performance. For example, in the
worst case (LDD on KR), omitting early exit may incur up to $3\times$ slowdown
compared to the full API. Omitting early exit
does not significantly affect performance in most cases: 65 cases incurred
negligible (less than $1.03\times$) slowdown, 27 cases incurred more than
$1.1\times$ slowdown, and 7 cases incurred more $2\times$ slowdown compared to
the full API. On the other hand, we found that the worst case ($k$-core on TW)
for omitting parallel maps resulted in $1.8\times$ slowdown compared to the full
API; parallel map is especially important in high-degree graphs such as
TW. However, the overall effect on performance from omitting parallel map was
even less than from omitting early exit: without parallel maps, 84 cases
incurred negligible (less than $1.03\times$) slowdown, 9 cases incurred more
than $1.1\times$ slowdown, and 0 cases incurred more than $2\times$ slowdown
compared to the full API.

Based on these results, we use all available API functionality besides
\texttt{parallel\_map\_early\_exit} in~\secreftwo{frameworks-eval}{algs-eval}.


\subsection{Comparison with other frameworks}\label{sec:frameworks-eval}

The goal of this section is to demonstrate that \apinameshort is a good basis
for our large-scale graph container evaluation by comparing it to several other
graph frameworks, including some frameworks that introduced graph containers
which we study in our benchmark.  Our objective in this section is to show that
graph containers that run algorithms with \apinameshort do not lose out on
performance compared to running with other frameworks.  Many of the comparisons
in this section are {\em not apples-to-apples} due to minor variations in
algorithm implementations.  However, it does show that \apinameshort is
competitive with existing work for end-to-end performance, and grounds the
results and lessons we obtain through \apinameshort in a high-performance
starting point.

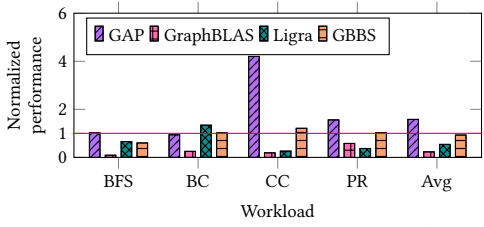
\begin{figure}
  \centering
  \begin{tikzpicture}
  \footnotesize
    \begin{axis}[
    width=7cm, height=3.5cm,
        ybar,
        legend pos=north west,
        ylabel={Normalized \\performance},
         ylabel style={align=center, text width=3cm},
        xlabel={Workload},
        xtick=data,
        bar width=4pt,
        legend columns=5,
        extra y ticks=1,
        enlarge x limits=0.14,
        symbolic x coords={BFS, BC, CC, PR, Avg},
        x tick label style={text width = 1cm, align = center},
        ymin = 0,
        ymax = 6
      ]

    \addplot [fill=safe-lavender, postaction={pattern=north east lines}] table [
          x=Algorithm, y=GAP, col sep=tab
        ] {tsvs/frameworks_bar.tsv};
        \addlegendentry{GAP}

         \addplot [fill=safe-pink, postaction={pattern=grid}] table [
          x=Algorithm, y=GraphBLAS, col sep=tab
        ] {tsvs/frameworks_bar.tsv};
        \addlegendentry{GraphBLAS}

            \addplot [fill=safe-teal, postaction={pattern=crosshatch}] table [
          x=Algorithm, y=Ligra, col sep=tab
          ] {tsvs/frameworks_bar.tsv};
          \addlegendentry{Ligra}

                      \addplot [fill=safe-peach, postaction={pattern=horizontal lines}] table [
          x=Algorithm, y=GBBS, col sep=tab
          ] {tsvs/frameworks_bar.tsv};
          \addlegendentry{GBBS}
          \addplot[red,sharp plot,update limits=false,] coordinates { ([normalized]-1,1) ([normalized]8,1) };
    \end{axis}
  \end{tikzpicture}
  \caption{Relative performance normalized to \apinameshort (up is good). A bar
    above 1 means there was speedup over \apinameshort.  Unlike the rest of this
    paper, this is not an apples-to-apples comparison because each system
    implements different algorithms and has been tuned for different graphs.
    This shows that BYO is competitive with the existing landscape of work,
    while still enabling flexibility in both container and algorithm choice. All
    systems besides GAP are frameworks, while GAP is a library of direct
    algorithm implementations. \new{All systems use CSR as the graph
      representation.}}
    \label{fig:frameworks-bar}
  \end{figure}

\figref{frameworks-bar} shows that \apinameshort achieves competitive
performance overall when compared with direct algorithm implementations from the
GAP benchmark suite~\cite{BeamAsPa15} as well as other high-performance
frameworks including Ligra, GraphBLAS, and GBBS.

On average, GAP supports algorithms about $1.6\times$ faster than \apinameshort,
but the majority of the performance gap comes from Connectivity (CC) because the
two systems implement very different algorithms for the problem.  GAP implements
the Afforest algorithm~\cite{SuttonBeBa18}, which is based on an idea similar to
direction-optimization that enables the algorithm to potentially examine far
fewer than all $m$ edges.  On the other hand, the GBBS algorithm we use is a
concurrent version of union-find which examines every edge, and is the
state-of-the-art for incremental connected components.  For the problems that
GAP and \apinameshort implement the same algorithms for, the average performance
gap narrows: GAP achieves about $1.15\times$ speedup over \apinameshort if we
exclude CC from consideration. These results demonstrate that the cost of
abstraction in \apinameshort is relatively small when compared to the direct
implementations from GAP.

The focus of this paper is on frameworks because they enable graph containers to
easily express a diverse set of algorithms. Direct implementations are
infeasible for large-scale evaluations because every algorithm must be
integrated with every data structure, combinatorially increasing the amount of
programming effort needed with every new data structure and algorithm.

In terms of frameworks, \apinameshort achieves very similar performance (within
about $1.05\times$) compared to GBBS, the starting point for \apinameshort's
implementation. Out of the frameworks we evaluated, Ligra/GBBS use similar
abstractions based on \vertexsubset/\edgemap~\cite{ShunBl13}. GBBS builds upon
Ligra with additional optimizations and functionality (e.g.,
bucketing~\cite{DhulBlSh17}). \apinameshort inherits these advancements from
GBBS, and both GBBS/\apinameshort achieve on average $1.7\times$ speedup over
Ligra. Finally, we also evaluate GraphBLAS~\cite{KepnerAaBa16, BulucMaMc17,
  Davis19, Davis23}, a state-of-the-art graph-algorithm framework based on
sparse linear algebra, and find that \apinameshort achieves about $4\times$
speedup over GraphBLAS. Our findings about GraphBLAS are consistent with a
recent comparison of GAP and GraphBLAS~\cite{Davis23} that shows that GraphBLAS
is slower than GAP because it cannot access optimizations such as kernel fusion.

\paragraph{Comparing to the original systems}
Furthermore, we verified that containers that run algorithms using \apinameshort
do not give up performance compared to their performance in the original systems
(with different frameworks for algorithms).  Specifically, when averaging across
algorithms and graphs, \apinameshort was between $1.1\times-1.8\times$ faster
than the original frameworks when integrated with PMA (both uncompressed and
compressed), SSTGraph, and CPAM.

\subsection{Comprehensive container evaluation}\label{sec:algs-eval}

\begin{table}
\scriptsize
\rev{
\begin{tabular}{@{}lrrrrrr@{}}
\toprule
\textit{Container} & \multicolumn{3}{c}{\textit{Slowdown over CSR}} &  \multicolumn{3}{c}{\textit{Bytes per edge}}  \\
                            &\textit{Average}  & \textit{95\%} & \textit{Max} &\textit{Min}  & \textit{Average} & \textit{Max}  \\
  \midrule
  \textit{\setapi (Vector of...)} \\
\midrule
\abslbtree & 1.26 & 1.9 & 2.3 \\
\abslbtree (inline) & \cellcolor{safe-green!25} 1.22 & 2 & 2.6 \\
\abslflathash & 1.40 & 2.3 & 3.4 \\
\abslflathash  (inline) & 1.29 & 2.1 & 2.6 \\
\stdset & 2.59 & 5.0 & 5.8 \\
\stdset (inline) & 2.37 & 4.9 & 5.6 \\
\stdunorderedset & 2.01 & 3.7 & 6.0 \\
\stdunorderedset (inline) & 1.90 & 3.5 & 5.9 \\
Aspen & \cellcolor{safe-green!25} 1.22 & 2 & 2.5 & 5.7&12.0&53.4 \\
Aspen (inline) & \cellcolor{safe-green!25} 1.14 & 1.7 & 2.0 &5.8&7.4&14.9 \\
Compressed Aspen & 1.44 & 2.1 & 2.6 &3.4&5.0&12.1 \\
Compressed Aspen (inline) & 1.34 & 1.9 & 2.6  &3.4&5.5&14.9 \\
CPAM & \cellcolor{safe-green!25} 1.16 & 1.4 & 1.5 &4.1&4.9&9.0 \\
CPAM (inline) & \cellcolor{safe-green!25} 1.11 & 1.5 & 1.6 &4.1&6.6&21.6 \\
Compressed CPAM & 1.37 & 1.7 & 1.9 &3.4&4.5&8.9 \\
Compressed CPAM (inline) & 1.30 & 1.8 & 2.1 &3.5&6.2&21.6 \\
PMA & \cellcolor{safe-green!25} 1.25 & 1.9 & 3.2 &8.1&13.9&46.5 \\
Compressed PMA & 1.35 & 1.9 & 3.3 &4.9&11.2&46.5   \\
Tinyset & 1.27 & 1.9 & 5.1 &5.5&8.6&26.5 \\
Vector & \cellcolor{safe-green!25} 1.07 & 1.4 & 1.9 &4.1&5.0&10.2 \\
  \midrule
  \textit{\containerapi} \\
\midrule
CSR & \cellcolor{safe-green!25} 1.00 & 1.0 & 1.0 &4.1&5.1&10.6 \\
Compressed CSR &  \cellcolor{safe-green!25} 1.23 & 1.5 & 1.6 &2.3&3.8&10.6 \\
DHB & \cellcolor{safe-green!25} 1.15 & 1.7 & 2.4 \\
PMA & \cellcolor{safe-green!25} 1.15 & 1.4 & 1.6 &10.0&12.3&24.2  \\
Compressed PMA & 1.31 & 2.0 & 2.2 &3.1&5.6&17.7 \\
SSTGraph & \cellcolor{safe-green!25} 1.25 & 1.5 & 2.4 &4.0&6.4&19.9 \\
Terrace & \cellcolor{safe-green!25} 1.20 & 2.0 & 3.3 &9.3&17.7&47.8 \\


\bottomrule
\end{tabular}
}
\caption{\rev{Data structure algorithm performance and space usage. All data structures are uncompressed unless otherwise
    specified.
    Each container's time is normalized to
    CSR's time averaged over all \numtotal settings of \numalgorithms algorithms
    $\times$ \numgraphs graphs. A number closer to 1 means better performance
    (higher is worse).  The 95\% and max columns show the 95th percentile and maximum
    slowdown over CSR across all algorithms and all graphs. 
    We also show the
    space usage of the different graph data structures in terms of bytes per edge.  
    } }
\label{tab:datastr-big-avg}
\end{table}


At a high level, the tested graph data structures are very similar on average,
but specialized data structures have an advantage over off-the-shelf structures
in terms of worst-case performance across problem
instances.~\tabref{datastr-big-avg} reports the average, 95th percentile, and
maximum slowdown over CSR for each data structure across all 100 problem
settings (\numalgorithms algorithms $\times$ \numgraphs graphs).

On average, we find that the overall difference between the best off-the-shelf
dynamic structure and the best specialized dynamic structure is within about
$1.1\times$. Specifically, the Abseil~\cite{absl} B-tree
combined with the inline optimization described in~\secref{byo-api}
incurs $1.22\times$ slowdown compared to CSR.
Furthermore, we find that the best specialized graph data structure on average
is a vector of uncompressed PaC-trees~\cite{DhulipalaBlGu22} + inline,
 which incurred $1.11\times$ slowdown relative to CSR.

The average differences between specialized structures are much smaller than
previously reported in other papers because \apinameshort standardizes the
evaluation and makes optimizations previously available in one system accessible
to all data structures. Specifically, we find that the specialized containers
(PaC-trees, Terrace, DHB, CPMA, SSTGraph and Aspen) incur between
$1.11-1.44\times$ slowdown on average relative to CSR.

These results do not invalidate previous evaluations because this paper compares
\emph{containers} directly rather than overall \emph{systems}. Previously,
papers that introduced containers were only able to compare their
systems (both the container and framework) because of the lack of a unified
easy-to-use framework. Therefore, previously-reported performance differences
were the result of variations in the framework as well as the container.  Additionally some existing work is designed for specific types of graphs or algorithms and thus tested on situations that they are expected to perform well. 

Although the off-the-shelf and specialized data structures achieve similar
performance on average, the specialized data structures have better overall
performance when looking at the holistic set of experiments.
\figref{how_close_structures} shows for how many experiment settings a given
data structure achieved within some slowdown relative to CSR. For example,
Abseil's B-tree with the inline optimization achieved within $1.25\times$ of
CSR's performance on 63 experiments, while PaC-trees with the inline
optimization achieved within $1.25\times$ of CSR's performance on 83
experiments.
\rev{In the worst case, Abseil's B-tree is $2.6\times$ slower than CSR on the
$k$-core algorithm on the KR graph due to a lack of parallel maps.  In contrast,
the vector of PaC-trees with the inline optimization, which is carefully
designed to be space-efficient and cache-friendly and can leverage parallel
maps, incurred only 
$1.16\times$ slowdown for the same problem and graph.} These results suggest that
specialized data structures can mitigate performance variations on challenging
instances such as high-degree graphs.

\para{Space usage}\mymarginpar{R4W4} \new{We also measure the space usage of all containers which support getting their memory usage in \tabref{datastr-big-avg}.  We find the bytes per edge varies significantly between graphs even when the container is fixed - by at least $2\times$ and sometimes up to $10\times$. In all cases, the worst-case bytes per edge is on the road graph due to its low degree. Finally, compressed data structures can reduce the space usage by $2\times$ compared their uncompressed counterparts.}

\begin{table*}
\scriptsize
\new{
\begin{tabular}{@{}l|rrrrrrrrrr@{}}
  \toprule
&  \textit{RD} & \textit{LJ}  & \textit{CO} & \textit{RM} & \textit{ER} &
                                                                         \textit{PR}
  & \textit{TW} & \textit{PA} & \textit{FS} & \textit{KR}\\
  \hline
   \textit{BFS} & \cellcolor{safe-teal!25}CPMA & \cellcolor{safe-pink!25}CPAM* & \cellcolor{safe-brown!25}Aspen* &  \cellcolor{safe-pink!25}CPAM* &  \cellcolor{safe-pink!25}CPAM* & \cellcolor{safe-yellow!25} TinySet &  \cellcolor{safe-pink!25}CPAM* &
                                                                            \cellcolor{safe-brown!25} Aspen*
                              & \cellcolor{safe-pink!25} CPAM &
                                                                \cellcolor{safe-teal!25}CPMA\\
    \textit{BC} &absl::FHS* &  \cellcolor{safe-pink!25} CPAM* & \cellcolor{safe-brown!25}Aspen* &  \cellcolor{safe-pink!25}CPAM* & \cellcolor{safe-pink!25} CPAM* &  \cellcolor{safe-plum!25}DHB&\cellcolor{safe-brown!25}Aspen* &\cellcolor{safe-brown!25} Aspen*
                              &\cellcolor{safe-brown!25}Aspen* &
                                                                 \cellcolor{safe-pink!25}CPAM*\\
    \textit{Spanner} &\cellcolor{safe-teal!25}PMA  &
                                                   \cellcolor{safe-pink!25}CPAM*
                              &  \cellcolor{safe-pink!25}CPAM*
                                            &\cellcolor{safe-brown!25}Aspen*  &
                                                                                \cellcolor{safe-pink!25}
                                                                                CPAM*
                                                                        &
                                                                          \cellcolor{safe-brown!25}Aspen* & \cellcolor{safe-plum!25} DHB & \cellcolor{safe-brown!25}Aspen*& \cellcolor{safe-plum!25} DHB &  \cellcolor{safe-plum!25}DHB \\
  \hline
    \textit{LDD} &\cellcolor{safe-teal!25}PMA & \cellcolor{safe-pink!25}CPAM* &
                                                                                \cellcolor{safe-teal!25}CPMA
                                            & \cellcolor{safe-pink!25} CPAM* &
                                                                               \cellcolor{safe-pink!25}
                                                                               CPAM*
                                                                        &
                                                                          \cellcolor{safe-plum!25} DHB &\cellcolor{safe-teal!25} CPMA &  \cellcolor{safe-pink!25}CPAM* & \cellcolor{safe-teal!25}CPMA& \cellcolor{safe-teal!25}CPMA \\
    \textit{CC} &absl::FHS* & \cellcolor{safe-brown!25}Aspen*
                              &\cellcolor{safe-brown!25} Aspen
                                            &\cellcolor{safe-brown!25} Aspen
                                                          &\cellcolor{safe-brown!25}
                                                            Aspen &
                                                                    \cellcolor{safe-brown!25}Aspen
  &\cellcolor{safe-brown!25} Aspen &  \cellcolor{safe-plum!25}DHB&
                                                                   \cellcolor{safe-plum!25}DHB
                                            & \cellcolor{safe-plum!25} DHB \\
  \hline
  \textit{ADS} &\cellcolor{safe-yellow!25}SSTGraph &\cellcolor{safe-yellow!25} TinySet &\cellcolor{safe-yellow!25}SSTGraph &\cellcolor{safe-green!25}Terrace & \cellcolor{safe-pink!25} CPAM & absl::btree &\cellcolor{safe-teal!25} PMA
  & \cellcolor{safe-plum!25} DHB& \cellcolor{safe-plum!25} DHB & \cellcolor{safe-plum!25} DHB\\
   \textit{KCore} &  \cellcolor{safe-pink!25}C-CPAM* &  \cellcolor{safe-pink!25}C-CPAM* & \cellcolor{safe-pink!25} CPAM & \cellcolor{safe-yellow!25}SSTGraph & \cellcolor{safe-yellow!25}SSTGraph &\cellcolor{safe-yellow!25} TinySet &
                                                                               \cellcolor{safe-pink!25}CPAM
                & \cellcolor{safe-pink!25} CPAM*&  \cellcolor{safe-pink!25}CPAM* &\cellcolor{safe-brown!25} Aspen*\\
\hline
   \textit{Coloring} &\cellcolor{safe-teal!25} PMA &\cellcolor{safe-teal!25}PMA &\cellcolor{safe-teal!25} PMA &\cellcolor{safe-teal!25} PMA &\cellcolor{safe-teal!25} PMA &\cellcolor{safe-teal!25} PMA &\cellcolor{safe-teal!25}PMA &\cellcolor{safe-teal!25} PMA&\cellcolor{safe-teal!25} PMA & \cellcolor{safe-teal!25}PMA(V) \\

  \textit{MIS} &\cellcolor{safe-teal!25}PMA & \cellcolor{safe-pink!25} CPAM* &
                                                                              \cellcolor{safe-yellow!25} TinySet & \cellcolor{safe-green!25}Terrace & \cellcolor{safe-pink!25} CPAM & \cellcolor{safe-yellow!25}TinySet& \cellcolor{safe-brown!25}Aspen* &  \cellcolor{safe-pink!25}CPAM*
                              &
                                \cellcolor{safe-brown!25}Aspen*&\cellcolor{safe-brown!25}Aspen*
  \\
  \hline
  \textit{PR} & \cellcolor{safe-plum!25}DHB &\cellcolor{safe-brown!25} Aspen* & \cellcolor{safe-brown!25}Aspen & \cellcolor{safe-green!25}Terrace &  \cellcolor{safe-pink!25}CPAM* & \cellcolor{safe-brown!25}Aspen&\cellcolor{safe-brown!25}Aspen*& \cellcolor{safe-yellow!25}TinySet &
                                                                                \cellcolor{safe-teal!25}PMA
                                                                                (V)
  &  \cellcolor{safe-plum!25}DHB\\

  \hline
\end{tabular}
}
\caption{The fastest container for every graph $\times$ algorithm combination. *
  next to a container denotes the \setapi version with the inline
  optimization. CPAM/C-CPAM refers to the uncompressed/compressed version of
  CPAM, respectively. PMA(V) refers to the vector of PMAs using the
  \setapi, and PMA refers to the single PMA
  using the \containerapi.}
\label{tab:winners}
\end{table*}


\new{\subsection*{Guidance for choosing  graph containers}}

\new{We first analyze how different specialized graph containers perform on
  different problem settings. Next, we compare the performance of different
  container configurations on the whole. Specifically, we analyze the effect of
  compression, the inline optimization (via the \setapi) and the effect of
  collocated data (via the \containerapi).}

\mymarginpar{R1D2} \mymarginpar{R4W2} \mymarginpar{R4D7}

\new{\paragraph{Relationship between containers and problem settings}
  ~\tabref{winners} shows the fastest container for each combination of
  graph and algorithm tested. These results provide guidance for choosing among
  containers for different graph and algorithm types.

  Overall, we find that the optimized tree-based containers (CPAM and Aspen)
  achieve the best performance most frequently on different problem
  settings. CPAM performs especially well on the ER graph - it is the fastest on
  7/10 algorithms. We conjecture that its performance is due to the uniform
  degree distribution and relatively high average degree in ER.

  Several other containers exhibit strengths in specific algorithm or graph
  categories:

  \begin{itemize}
  \item The PMA is the fastest container on the \emph{Coloring} algorithm for
    all graphs. Coloring is a covering-type algorithm that requires iterating
    over the entire graph in any order, which the PMA is well-suited
    due to being optimized for contiguous memory access.
  \item DHB achieves the best performance more often on \emph{large graphs}
    (i.e., on PA, FS, and KR). We conjecture that DHB is well-suited to large
    graphs because it uses custom memory allocations that enable it to
    store more data contiguously.
  \item Terrace has the best performance on some algorithms (ADS, MIS, and PR)
    when run on the RMAT graph. Terrace is optimized for skewed graphs, and RMAT
    is a synthetic skewed graph.

   \item On very sparse graphs, e.g., RD, data structures with fewer pointers
   and co-located memory such as PMA, CPMA, and SSTGraph are the best choice due to
   the improved locality of these data structures when the average degree of the graph
   is extremely low.

  \end{itemize}

  To summarize, CPAM and Aspen are solid choices for overall performance, but if
  a user has a specific algorithm or graph class that they are optimizing for,
  the trends noted above can help them select a different container.
}


Next, we compare classes of data
structures and the optimizations possible through the different APIs in
\apinameshort.

\paragraph{Effect of compression on containers}
These results demonstrate that compression does not help performance on the
tested data structures and graphs. Specifically, the compressed versions of CSR,
Aspen, PaC-trees, and PMA incurred between $1.1-1.23\times$ additional slowdown
over CSR compared to the uncompressed versions.

These results stand in contrast to previous work that demonstrated speedups due
to compression~\cite{DhulipalaBlSh19,WheatmanBuBu23, ShunGhBl15,
  DhulipalaBlGu22, DhulBlSh21} because \apinameshort provides a highly-optimized
framework that improves upon the frameworks in prior comparisons. Although the
algorithms are still memory bound, the computational overhead from decompression
impacts performance in algorithms in \apinameshort because margins for
optimization and overall improvement are smaller, so any additional work has a
more pronounced effect on performance.

We focus our evaluation in this paper on graphs with up to billions of edges to
match commonly-used dataset sizes specified by graph benchmark suites such as
GAP~\cite{BeamAsPa15} and LDBC Graphalytics~\cite{IosupHeNg16}. However,
compression is important for feasibility as well as performance as graphs scale
even larger to hundreds of billions of edges. Without compression, these graphs
cannot fit in memory.

\paragraph{Effects of inline optimization in \setapi}
~\tabref{datastr-big-avg} demonstrates that the inline optimization improves
performance of set data structures on average by \inlineavgspeedup (between
$1.03-1.09\times$ averaged across both graphs and algorithms).  The inline
optimization was introduced in the Terrace~\cite{PandeyWhXu21} graph system as
part of its hierarchical data structure design. However, we incorporate this
optimization into the \setapi in \apinameshort to benefit all data structures.

\paragraph{Advantages of \containerapi}
Finally, we find that collocating data between sets (enabled by the
\containerapi described in~\secref{graph-api}) improves performance by between
$1.02-1.1\times$ because it reduces indirections between neighbor lists.
For example, as shown in~\tabref{datastr-big-avg}, the vector of vectors (the
\setapi version of CSR) incurs $1.07\times$ slowdown over CSR. Similarly, the
single PMA is faster on average compared to the vector of PMAs (one per vertex),
and SSTGraph is faster on average compared to the vector of tinyset (the \setapi
version of SSTGraph). These results demonstrate that collocating the data
improves performance on a diverse set of algorithms and graphs.

\subsection{Batch-insert evaluation}\label{sec:batch-eval}

We measure batch-insert throughput, an important feature of dynamic-graph
containers, and report the results in \figref{inserts_with_sort}.

\paragraph{Setup}
To evaluate update throughput, we first insert all edges from the TW graph (a
billion-edge scale graph) into the system under test. We then insert a new batch
of directed edges (with potential duplicates) to the existing graph in the
system under test and then delete that batch of edges from the graph. We repeat
this procedure 3 times per batch to generate multiple trials. To generate edges
for inserts/deletes, we sample directed edges from an \rmat
generator~\cite{ChakZhFa04} (with $a = 0.5; b = c = 0.1; d = 0.3$, matching the
distribution from prior work on dynamic-graph systems~\cite{PandeyWhXu21,
  DhulipalaBlGu22, WheatmanBuBu23}).

\paragraph{Discussion}
\rev{ The off-the-shelf data structures exhibit the classical tradeoff between
  algorithm and update performance. The fastest off-the-shelf container for
  inserts is Abseil's flat hash set, which achieved between $3\times$ speedup
  over Abseil's B-tree on inserts for large batches. On the other hand, Abseil's
  flat hash set incurred more slowdown on algorithms over CSR compared to
  Abseil's B-tree, as shown in~\figref{how_close_structures}
  and~\tabref{datastr-big-avg}. We plot the set container variants with the
  inline optimization because we found they did not impact the insertion
  throughput greatly.

  Specialized structures can overcome the algorithm-update tradeoff on the
  largest batches with special support for batch inserts. For example, the
  batch-parallel PMA (using the \containerapi) achieved $2-5\times$ speedup over
  the Abseil B-tree on the larger batches due to the PMA's native support for
  parallel batch inserts. Furthermore, we found that DHB, a hash-based data
  structure that uses the \containerapi, achieves $1-3\times$ speedup over the
  B-tree on the batch insertions. Both the PMA and DHB achieve better overall
  performance on algorithms compared to the B-tree (\tabref{datastr-big-avg}).

  Other specialized structures may trade update speed for algorithm performance
  compared to the B-tree. For example, CPAM (with inline) incurred $1.15\times$
  slowdown compared to the B-tree (with inline) on batch inserts when averaging
  across batch sizes.
  Furthermore, we found that Terrace was about $10\times$ slower than the B-tree
  for batch inserts on average across batches because Terrace maintains many
  edges in a concurrent contiguous PMA and therefore cannot trivially
  parallelize across vertices.  }

These results suggest there is potential in developing specialized
batch-parallel data structures that overcome the tradeoff for batch inserts
without giving up on algorithm performance.
\new{The results also show the importance of how the data needs to be processed before it can actually be inserted into the data structures.  The fastest approaches for large batches are those that only require a semi-sort.  The fastest for mid sized batches and second fastest for large batches require a full sort, but that perform merges for the data.}

\section{Conclusion}
This paper introduces \apinameshort, an easy-to-use, high-performance, and expressive
graph-algorithm framework.
\apinameshort enables apples-to-apples comparisons between dynamic-graph containers by decoupling the graph containers from algorithm implementations.
The \apinameshort interface 
is simple, 
enabling comprehensive comparisons of new containers on a diverse set of applications with minimal programming effort.

We have conducted a large-scale evaluation of 27 graph containers using \apinameshort to express algorithms.
The results demonstrate that the differences between graph containers are smaller than what is commonly reported in papers introducing new graph containers.
We attribute this discrepancy to the fact that these papers often perform end-to-end comparisons between graph systems, which vary both the framework and the container.
Moreover, our results demonstrate that while on average off-the-shelf data structures achieve highly competitive performance with specialized data structures, they leave significant performance on the table for certain algorithms/graphs. 

We believe that \apinameshort will spark exploration into specialized data structures for specific cases where off-the-shelf data structures do not do well, as well as on data structures that support fast updates without giving up on algorithm
performance.


\begin{acks}
This work is supported by NSF grants CCF-2103483, CCF-2238358, CCF-2339310, CNS-2317194, IIS-2227669, and OAC-2339521.
We thank the anonymous reviewers for their useful comments.

\end{acks}

\clearpage
\balance
\bibliographystyle{ACM-Reference-Format}
\bibliography{sample}

\end{document}